\documentclass[journal]{IEEEtran}

\usepackage{cite}
\usepackage{amsmath,amssymb,amsfonts,amsthm}
\usepackage{algorithm}
\usepackage{algorithmic}
\usepackage{graphicx}
\usepackage{textcomp}
\usepackage{xcolor}
\usepackage{mathtools}
\usepackage{graphicx}
\usepackage{booktabs}
\usepackage{makecell}
\usepackage[font={small}]{caption}
\usepackage{subcaption}
\usepackage{cite}
\usepackage{breqn}
\usepackage{mathrsfs}
\usepackage{accents}
\usepackage{acronym}
\usepackage{soul}
\usepackage{bm}
\usepackage{url}
\usepackage{wrapfig}
\usepackage{todonotes}
\usepackage{verbatim}
\usepackage{tikz}
\usepackage{pdfpages}
\usepackage{multirow}
\usepackage{fancyhdr}

\DeclareMathOperator*{\argmin}{arg\,min}
\DeclareMathOperator*{\argmax}{arg\,max}

\graphicspath{{./figures/}}
\newcommand{\eq}[1]{Eq.~\eqref{#1}}
\newcommand{\fig}[1]{Fig.~\ref{#1}}
\newcommand{\tab}[1]{Tab.~\ref{#1}}
\newcommand{\secref}[1]{Section~\ref{#1}}
\newcommand{\alg}[1]{Alg.~\ref{#1}}

\newcommand{\tabentry}[1]{\footnotesize{{#1}}}

\newcommand{\mytexttilde}{{\raise.17ex\hbox{$\scriptstyle\mathtt{\sim}$}}}

\begin{document}

\title{milliTRACE-IR: Contact Tracing and Temperature Screening via mm-Wave and Infrared Sensing}
\author{Marco Canil$^\ddag$, \IEEEmembership{Graduate Student Member, IEEE}, Jacopo Pegoraro$^\ddag$, \IEEEmembership{Graduate Student Member, IEEE} and Michele Rossi, \IEEEmembership{Senior Member, IEEE} \\
\IEEEauthorblockA{\small{$^\ddag$ These authors contributed equally to this research. $^*$ Corresponding author email \texttt{pegoraroja@dei.unipd.it}. The authors are with the Department of Information Engineering at the University of Padova, via Gradenigo 36/b, 35131, Padova, Italy}}}


\maketitle
\thispagestyle{fancy}

\begin{abstract}
Social distancing and temperature screening have been widely employed to counteract the COVID-19 pandemic, sparking great interest from academia, industry and public administrations worldwide. 
While most solutions have dealt with these aspects separately, their combination would greatly benefit the continuous monitoring of public spaces and help trigger effective countermeasures.
This work presents milliTRACE-IR, a joint mmWave radar and infrared imaging sensing system performing unobtrusive and privacy preserving human body temperature screening and contact tracing in indoor spaces. milliTRACE-IR combines, via a robust sensor fusion approach, mmWave radars and infrared thermal cameras. 
It achieves fully automated measurement of distancing and body temperature, by jointly tracking the subjects's faces in the thermal camera image plane and the human motion in the radar reference system. 
Moreover, milliTRACE-IR performs contact tracing: a person with high body temperature is reliably detected by the thermal camera sensor and subsequently traced across a large indoor area in a non-invasive way by the radars. When entering a new room, a subject is re-identified among several other individuals by computing gait-related features from the radar reflections through a deep neural network and using a weighted extreme learning machine as the final re-identification tool.
Experimental results, obtained from a real implementation of milliTRACE-IR, demonstrate decimeter-level accuracy in distance/trajectory estimation, inter-personal distance estimation (effective for subjects getting as close as $\boldsymbol{0.2}$~m), and accurate temperature monitoring (max. errors of $\boldsymbol{0.5}$~°C). Furthermore, milliTRACE-IR provides contact tracing through highly accurate ($\boldsymbol{95\%}$) person re-identification, in less than $\boldsymbol{20}$ seconds.
\end{abstract}


\IEEEpeerreviewmaketitle

\section{Introduction}
\label{sec:introduction}

This work tackles the problem of designing a real-time, integrated radio and infrared sensing system to jointly perform unobtrusive elevated skin temperature screening and privacy preserving contact tracing in indoor environments.

Lately, \emph{social distancing} has become a primary strategy to counteract the \mbox{COVID-19} infection. Many research works~\cite{nguyen2020enabling, thu2020effect} have shown that it is an effective non-pharmacological approach and an important inhibitor for limiting the transmission of many contagious diseases such as H1N1, SARS, and \mbox{COVID-19}. Along with social distancing, \emph{elevated skin temperature detection} and \emph{contact tracing} have proven to be key to effectively contain the pandemic~\cite{de2021lessons}. However, available methods to enforce these countermeasures often rely on RGB cameras and/or apps that need to be installed and continuously run on people's smartphones, often rising privacy concerns~\cite{ali2021study}. Moreover, currently adopted methods to screen people's temperature require individuals to stand in front of a thermal sensor, which may be impractical in heavily frequented public places. 

\begin{figure}[t!]
	\begin{center}   
		\includegraphics[width=8.1cm]{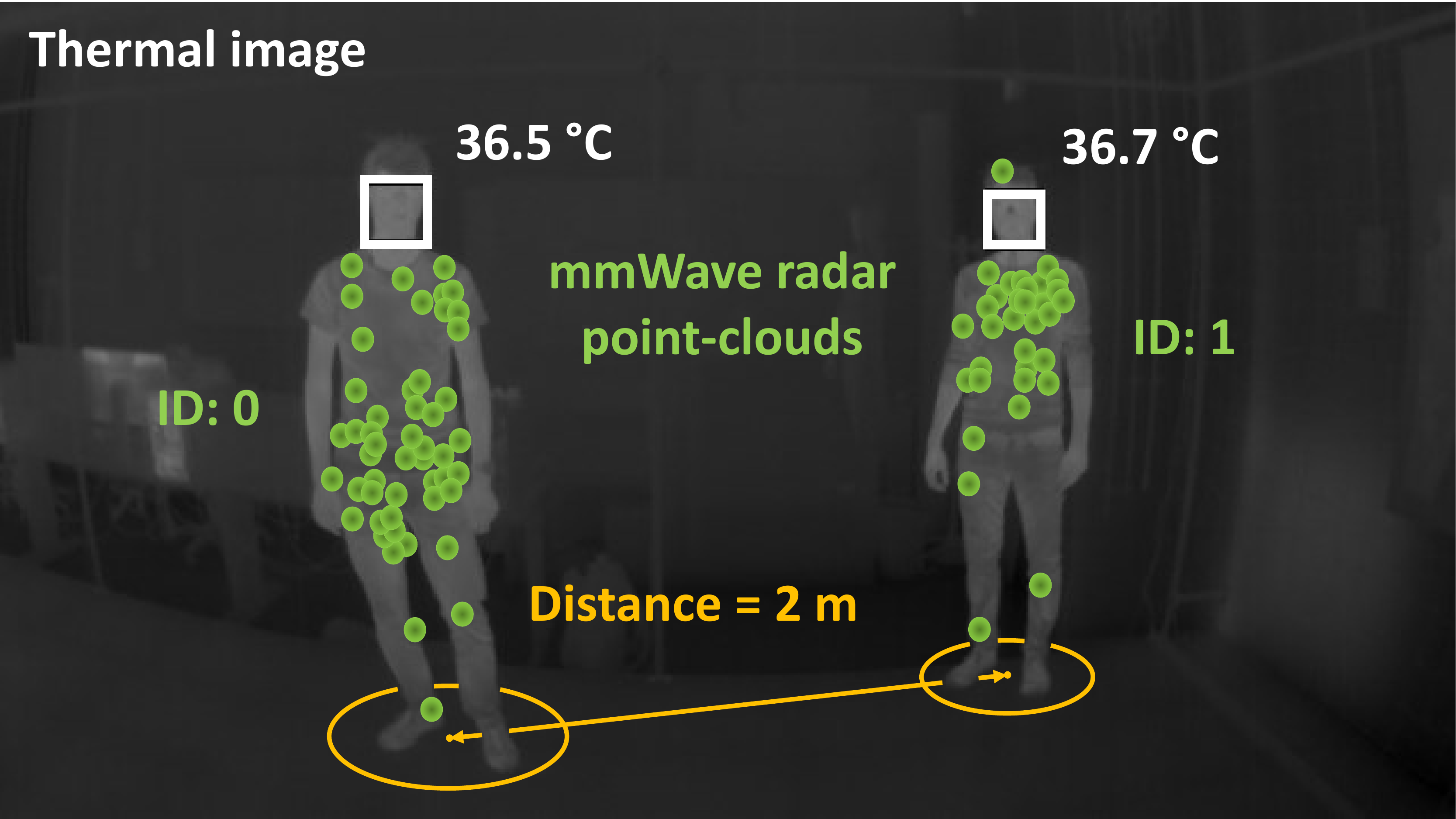} 
		\caption{milliTRACE-IR performs body temperature screening and interpersonal distance estimation via sensor fusion of an infra-red thermal camera and mmWave radars. Individual gait features contained in the mmWave reflections enable contact tracing across different rooms.}
		\label{fig:g-abstr}
	\end{center}
\end{figure}

Here, milliTRACE-IR, a joint mmWave radar and infrared imaging sensing system is designed and validated. milliTRACE-IR performs unobtrusive and privacy preserving human body temperature screening and contact tracing in indoor spaces (see \fig{fig:g-abstr}). Next, its main components are discussed, emphasizing their novel aspects and the joint processing of the acquired sensor data.    

\noindent\textbf{mmWave radar:} The radar analyzes the reflections of a transmitted mmWave signal off the individuals that move in the monitored environment, returning \emph{sparse point-clouds} that carry information about the subjects' locations and the velocity of their body parts. A novel point-cloud clustering method is designed, combining Gaussian mixtures~\cite{bishop2006pattern} and the density-based DBSCAN~\cite{ester1996density} algorithm, to distinguish the mmWave radio reflections from the subjects, as they move as close as $0.2$~m to one another. The so obtained point-cloud clusters are used to track the subjects' positions in the physical space by means of a Kalman Filter (KF)~\cite{kalman1960new}, and to obtain their gait-related features through a deep-learning based feature extractor. Finally, a novel person re-identification algorithm is proposed by exploiting weighted extreme learning machines (WELM).

\noindent\textbf{Thermal camera:} The infrared imaging system, or thermal camera (TC), returns images whose pixels contain information on the \emph{temperature of the objects} in the TC field of view (FoV). To measure the subjects' temperature, at first, YOLOv3~\cite{farhadi2018yolov3} is used to perform face detection in the TC images, by bounding those areas containing a human face. Hence, the obtained bounding boxes are tracked through an Extended Kalman Filter (EKF)~\cite{ribeiro2004kalman} and the subjects' temperature is estimated by accumulating readings for each EKF track, according to a dedicated estimation and correction procedure. Through the EKF, the subject's distance from the TC is also estimated from the size of the corresponding bounding box by considering the non-linear part of the EKF, which is approximated by fitting a function over a set of experimental data points.

\noindent\textbf{Radar and thermal camera data fusion:} Tracks in the radar reference systems are associated with those in the TC image plane via an original algorithm that finds optimal matches for the readings taken by the two sensors, through their {\it joint} analysis. This makes it possible to take temperature measurements from a subject and reliably associate them with the highly precise tracking of his/her movement performed by the radar. In addition, the joint analysis of radar and TC data allows refining the temperature estimated through the TC: to mitigate the influence of the distance on the temperature readings~\cite{savazzi2020processing}, a regression function that provides temperature correction coefficients is fit from training data. The final temperatures are obtained using such function with the accurate distances retrieved from the radar.


Hence, once a subject's temperature is measured, it is associated with the corresponding radar track and the subject's movements and contacts inside the building are accurately monitored, by re-identifying the subject as he/she moves across the FoV of different radar devices. To the best of the author's knowledge, milliTRACE-IR is the first system that achieves temperature screening and human tracking through the joint analysis of radar and TC signals. Furthermore, it concurrently performs body temperature screening and contact tracing, while these aspects have been previously dealt with separately. A sensible usage model for the system is as follows: the TCs shall be deployed in strategic locations to allow an effective temperature screening, such as facing the building/room entrance, to ensure that people's faces are seen frontally for a reasonable amount of time, and that their TC images are only taken when they enter or leave the building/room. On the other hand, the radar can be utilized to track the subjects while moving inside the monitored indoor space. This ensures higher privacy with respect to RGB cameras.

The main contributions of the present work are:
\begin{enumerate}
	\item milliTRACE-IR, a joint {\it \mbox{mmWave} radar and infrared imaging sensing system} that performs unobtrusive and privacy preserving human body temperature screening and contact tracing in indoor spaces is designed and validated through an extensive experimental campaign.
	\item A novel {\it data association} method is put forward to robustly associate tracks obtained from the \mbox{mmWave} radar and from the TC, where the radar returns the people coordinates in the physical space and the TC identifies people's faces in the thermal image space. The achieved precision and recall in the associations are as high as $97\%$.
	\item An original {\it clustering algorithm for mmWave point-clouds} is devised, making it possible to resolve the radar reflections from subjects as close as $0.2$~m.
	\item A new WELM based {\it person re-identification} procedure is presented. The WELM is trained at runtime on previously unseen subjects, achieving an accuracy of $95\%$ over six subjects with only $3$ minutes of training data.
	\item A novel method is designed to perform {\it elevated skin temperature screening} as people move freely within the FoV of the TC, without requiring them to stop and stand in front of the thermal sensor. For this, a dedicated approach is presented to mitigate the distortion in the TC temperature readings as a function of the distance, by also leveraging the accurate distance measures from the radar. Through this method, worst-case errors of $0.5$~°C are obtained.
\end{enumerate}
The paper is organized as follows. In \secref{sec:rel-work}, the related work is discussed. \secref{sec:prel} introduces some basic concepts about mmWave radars and thermal imaging systems, while in \secref{sec:approach} the proposed approach is thoroughly presented. In \secref{sec:implementation}, the implementation of milliTRACE-IR is described, while \secref{sec:results} contains an in depth evaluation of milliTRACE-IR on a real experimentation setup. Concluding remarks are provided in \secref{sec:conclusion}.

\section{Related Work}
\label{sec:rel-work}

In the literature, almost no work has focused on a joint approach to social distancing and people's body temperature monitoring which preserves the privacy of the users. Here, several prior works in related areas are discussed, highlighting the differences with respect to the proposed system.

\noindent \textbf{Social distancing monitoring:} Social distancing has been one of the most widely employed countermeasures to contagious diseases outbreaks~\cite{nguyen2020enabling}. Real-time monitoring of the distance between people in workplaces or public buildings is key for risk assessment and to prevent the formation of crowds. Existing approaches use either wireless technology like Bluetooth or WiFi~\cite{faragher2015location, galvan2012wifi}, which require the users to carry a mobile device, or camera-based systems~\cite{cristani2020visual}, which are privacy invasive.
Other approaches use the received signal strength indication (RSSI) from cellular communication protocols~\cite{nguyen2020enabling} or wearables~\cite{bian2020wearable}, although these are often inaccurate, especially when used in crowded places~\cite{nguyen2020enabling}. A lot of effort has been put into designing person detection and tracking algorithms for crowd monitoring and people counting~\cite{rezaei2020deepsocial} by using fixed surveillance cameras and mobile robots~\cite{sathyamoorthy2020covid}. The main drawbacks of these methods are the intrinsic difficulty in estimating the distance between people from images or videos, along with the fact that the users have to be continuously filmed during their daily lives, which raises privacy concerns.

Concurrently, a large body of work has focused on ultra-wideband (UWB) transmission for people tracking~\cite{knudde2017indoor,zhao2019mid}, e.g., using mmWave radars, as these naturally allow measuring distances with decimeter-level accuracy. However, none of these works has tackled the problem of estimating interpersonal distances when people are very close to one another for extended periods of time; this is especially difficult with radio signals, as the separation of the reflections from different subjects becomes challenging. 

\noindent \textbf{Passive temperature screening:} Infrared thermography is widely adopted for non-contact temperature screening of people in public places~\cite{dell2021noncontact}. Due to the COVID-19 pandemic, there has been a growing interest in developing screening methods to measure the temperature of multiple subjects simultaneously, without requiring them to collaborate and/or to carry dedicated devices~\cite{ferrari2021inner}. Approaches that involve the use of RGB cameras, e.g.,~\cite{lewicki2020ai}, share the aforementioned privacy-related limitations.

The authors of~\cite{savazzi2020processing} developed a Bayesian framework to measure the body temperature of multiple users using low-cost passive infrared sensors. The distance from the sensors and the number of subjects is also obtained. However, the working range of this system is very short (around $1.5$~m for precise temperature estimation), so it is deemed unapt for monitoring a large indoor area.

\noindent \textbf{Radar-thermal imaging association and fusion:} Sensor fusion between radars and RGB cameras has been extensively investigated, see, e.g.,~\cite{zhang2019extending,nobis2019deep}, while the joint processing of mmWave radar data and infrared thermal images was marginally treated~\cite{ulrich2018person}. In addition, the last paper only deals with the detection of humans using thermal imaging and does not address body temperature screening.

The present work is focused on the \textit{data association} between a thermal camera and a mmWave radar over short periods of time, using the accurate radar distance estimates to refine the temperature reading. This makes it possible to consider scenarios where the thermal camera only covers a small portion of the environment (e.g., the entrance) so as to preserve the subjects' privacy, while a mmWave radar network can effectively monitor the whole indoor space. 

\noindent \textbf{mmWave radar person re-identification (Re-Id):} Radio-frequency (RF) based person Re-Id is a recent research topic. So far, many works have focused on person identification~\cite{vandersmissen2018indoor, meng2020gait}, where the subjects to identify have been previously seen by the system, typically via a preliminary training phase. Re-Id is more challenging, as it addresses the recognition of \textit{unseen} subjects, for which only a few radio samples are collected during system operation. 
Differently from camera image based Re-Id methods~\cite{ye2021deep}, RF approaches need to profile the users across time intervals of a few seconds, to extract robust person specific features~\cite{fan2020learning}. To the best of the author's knowledge, only two works have proposed solutions to this problems~\cite{fan2020learning, cheng2021person}. In both cases, a deep learning method trained on a large set of users is used to extract features from the human gait. At test time, the features obtained from the subjects to be re-identified are compared against those of a set of known individuals using distance-based similarity scores.  
This approach entirely depends on the feature extraction process, and the classifier does not learn to refine its decisions at {\it runtime}, as new samples become available. This is a weakness, as the gait features extracted from mmWave radars are known to be variable, e.g., across different days~\cite{pegoraro2021realtime}. Conversely, milliTRACE-IR combines deep feature extraction with fast classifiers which are continuously trained and refined as new data is collected; this improves the robustness of the identification task.

\section{Preliminaries}
\label{sec:prel}

In this section the main working principles of the sensing technologies used in this work are summarized, namely, \textit{frequency-modulated continuous wave} (FMCW) mmWave radars and infrared thermal cameras.

\subsection{mmWave FMCW Radar}
\label{sec:fmcw-radar}

A MIMO FMCW radar allows the joint estimation of the distance, the radial velocity and the angular position of the targets with respect to the radar device~\cite{patole2017automotive}. It works by transmitting sequences of \textit{chirp} signals, linearly sweeping a bandwidth $B$, and analyzing their copies, which are reflected back from the environment. A full chirp sequence, termed \textit{radar frame}, is repeated with period $\Delta$ seconds.

\subsubsection{Distance, velocity and angle estimation} 
By computing the frequency shift induced by the  delay of each reflection, the radar allows obtaining the distance and velocity of the targets with high accuracy. 
The use of multiple receiving antennas, organized in a \textit{planar array}, allows obtaining the angle-of-arrival (AoA) of the reflections along the azimuth and the elevation dimensions, leveraging the different frequency shifts measured by the different antenna elements. This enables the localization of the targets in the physical space.

\begin{figure*}[t!]
	\begin{center}   
		\includegraphics[width=12.7cm]{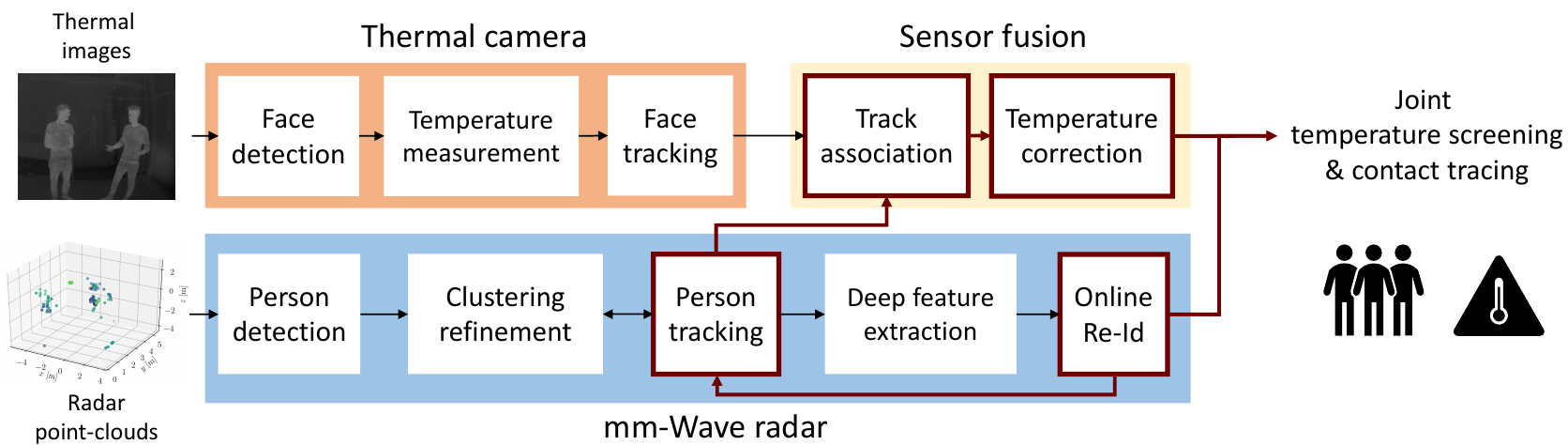} 
		\caption{milliTRACE-IR signal processing workflow.}
		\label{fig:workflow}
	\end{center}
	\vspace{-0.5cm}
\end{figure*}
\subsubsection{Radar detection} The raw output of the radar is typically high dimensional for mmWave devices, due to the high resolution. To sparsify the signal and perform a detection of the main reflecting points, a typical approach is the \textit{constant false alarm rate} (CFAR) algorithm~\cite{richards2010principles}, which consists of applying a dynamic threshold on the power spectrum of the output signal. 
A further processing step is required to remove the reflections from static objects, i.e., the \emph{clutter}. This operation is performed using a \textit{moving target indication} (MTI) high pass filter that removes the reflections with Doppler frequency values close to zero~\cite{richards2010principles}.

\subsubsection{Radar point-clouds} After the detection phase, a human presence in the environment typically generates a large number of detected points. This set of points, usually termed radar \textit{point-cloud}, can be transformed into the $3$-dimensional Cartesian space ($x-y-z$) using the distance, azimuth and elevation angles information of the multiple body parts. In addition, the velocity of each point is also retrieved, along with the strength of the corresponding signal reflection. 

In the following, the point-cloud outputted by the radar at frame $k$ is referred to as $\mathcal{P}_k$, containing a variable number of reflecting points. Each point, $\boldsymbol{p} \in \mathcal{P}_k$, is described by vector \mbox{$\boldsymbol{p} = \left[x, y, z, v, P^{\mathrm{RX}}\right]^T$}, including its coordinates $x,y,z$, its velocity $v$ and reflected power $P^{\mathrm{RX}}$.

\subsection{Infrared Thermal Cameras}
\label{sec:TC_preliminaries}

Infrared thermal imaging deals with detecting radiation in the long-infrared range of the electromagnetic spectrum \mbox{($\sim8-15$~$\mu$m)} and producing images of that radiation, called {\it thermograms}. According to the \emph{Planck's Law}, infrared radiation is emitted by all objects with temperature $T > 0$~K~\cite{DIASInfraredSystems}.
Since the radiation energy emitted by an object is positively correlated to its temperature, from the analysis of the received radiation it is possible to measure the object's temperature.

A thermographic camera, or \emph{thermal camera}, is a device that is capable of creating images of the detected infrared radiation. The operating principle is quite similar to that of a standard camera, and the same relations described by the so-called \emph{pinhole camera model} hold~\cite{princeCVMLI2012}. Within this approximation, the coordinates of a point $\boldsymbol{a}=[a_x,a_y,a_z]^T$ in the three-dimensional space are projected onto the image plane of an ideal pinhole camera through a very small aperture. Mathematically, this operation is described as $\boldsymbol{a}^{\rm proj} = \boldsymbol{\Psi}\boldsymbol{a}$, where $\boldsymbol{a}^{\rm proj}$ is the projected point and $\boldsymbol{\Psi}$ is the intrinsic matrix of the camera that contains information about its focal lengths, pixel dimensions and position of the image plane. 
However, when dealing with a real thermal camera, this approximation may be insufficient and the \emph{radial} and/or \emph{tangential} distortions introduced by the use of a lens and by inaccuracies in the manufacturing process may additionally have to be accounted for. On the image plane, an array of infrared detectors is responsible for measuring the received radiation, which is sampled and quantized to produce a digital information. The pixels of the final image that is returned by a thermal camera contain information about the temperature of the corresponding body/object part, encoded into the pixel intensity.


\section{Proposed Approach}
\label{sec:approach}

This work considers the problem of monitoring an indoor environment covered by multiple mmWave radar sensors, which span over different rooms and corridors. A few infrared thermal cameras are placed at strategic locations to perform accurate temperature screening of the people in the indoor space without compromising their privacy, e.g., at the building's entrance.

From a high-level perspective, milliTRACE-IR performs the following operations.

\noindent $(1)$ \textbf{Person detection and temperature measurement:} When people enter the monitored indoor space, the system concurrently performs face detection from the infrared images captured by the thermal camera and person detection using the mmWave radar point clouds. 
\begin{enumerate}
	\item From the thermal camera (TC) images, a face detector is used to obtain bounding boxes enclosing the faces of the detected subjects, (\secref{sec:tc-ekf}). A measure of their body temperature is obtained from the intensity of the thermal image pixels in the bounding box, see \secref{sec:temp-est}. While milliTRACE-IR works independently of the specific face detector architecture used, in the implementation YOLOv3 is used~\cite{farhadi2018yolov3}.
	\item Concurrently, radar signal processing is used to detect and group the point-clouds from different subjects and estimate their positions (\secref{sec:mmwave-track}). A novel clustering algorithm based on DBSCAN and Gaussian Mixture models is put forward to separate the contributions of closeby subjects (\secref{sec:cluster-res}).
\end{enumerate}

\noindent $(2)$ \textbf{Radar-TC person tracking:} Kalman filtering (KF) is independently applied to the TC images and to the radar point-clouds to respectively track the subjects' movements within the thermal images and in the indoor Cartesian space. Standard KF-based tracking in the thermal image plane is here modified to achieve a coarse estimation of the distance of the subjects, based on the dimension of their face bounding box \secref{sec:tc-ekf}. In this phase, each subject track is associated with a unique numerical identifier.

\noindent $(3)$ \textbf{Radar-TC track association:} As a subject exits the FoV of the TC, his/her body temperature is associated with the corresponding trajectory from the mmWave radar, by performing a track-to-track association between TC tracks and radar tracks. This association algorithm is based on the subjects' distances from the TC, and on the radar estimated positions of the subjects, projected onto the thermal image plane (\secref{sec:data-assoc}). After the association, the temperature measurement is corrected accounting for the distance of each person from the TC, using the more precise distance estimates provided by the radar, \secref{sec:temp-est}.

\noindent $(4)$ \textbf{Radar-based person re-identification:} During the radar tracking process, the point-cloud sequences generated by each subject are collected and fed to a deep neural network that performs gait feature extraction (\secref{sec:dl-fextr}). The resulting gait features are organized into a labeled training set, where labels are obtained from the track identifiers. When a subject exits the FoV of a radar and enters that of another radar placed in a different room or corridor, a weighted extreme learning machine (WELM) based classifier~\cite{zong2013weighted} is trained on-the-fly and used to re-identify the subject at runtime (\secref{sec:elm-reid}). This robust and lightweight person Re-Id process, based on the gait features extracted from the radar point-clouds, enables contact tracing across large indoor environments.

\subsection{Notation}
\label{sec:notation}

The system operates at discrete time-steps, \mbox{$k=1,2, \dots $}, each with fixed duration of $\Delta$ seconds, also referred to as \textit{frame} in the following.  Boldface, capital letters refer to matrices, e.g., $\boldsymbol{X}$, with elements $X_{ij}$, whereas boldface lowercase letters refer to vectors, e.g., $\boldsymbol{x}$. 
$\boldsymbol{X}^{-1}$ denotes the inverse of matrix $\boldsymbol{X}$, and $\boldsymbol{x}^T$ denotes the transpose of vector $\boldsymbol{x}$. $\boldsymbol{x}_k$ refers to vector $\boldsymbol{x}$ at time $k$, $x_j$ refers to element $j$ of $\boldsymbol{x}$ and $(\boldsymbol{x}_k)_j$ is element $j$ of $\boldsymbol{x}_k$. $\mathcal{N}(\mu, \sigma^2)$ indicates a Gaussian random variable with mean $\mu$ and variance $\sigma^2$. Notation $||\boldsymbol{x}||_2$ indicates the Euclidean norm of vector $\boldsymbol{x}$, while $||\boldsymbol{x}||_{\boldsymbol{\Gamma}} = \sqrt{\boldsymbol{x}^T\boldsymbol{\Gamma} \boldsymbol{x}}$ denotes the norm induced by matrix $\boldsymbol{\Gamma}$. The diagonal matrix with elements $x_1, x_2, \dots, x_n$ is denoted by $\mathrm{diag}\left[x_1, x_2, \dots, x_n\right]$. $|\mathcal{X}|$ indicates the cardinality of set $\mathcal{X}$ while $\log (\cdot)$ denotes the natural logarithm.


\subsection{Thermal Camera: Face Detection and Tracking}
\label{sec:tc-ekf}

The detection of the subjects in the thermal camera images is performed by means of a face detector that computes rectangular bounding boxes delimiting the faces of the people within the FoV. The bounding boxes are used to track the positions of the subjects in the subsequent instants and to identify a region of interest (ROI) from which the temperature of the targets is obtained. milliTRACE-IR is independent of the particular face detector used, provided that it outputs bounding boxes enclosing the faces of the subjects. In the implementation, YOLOv3~\cite{farhadi2018yolov3} is used due to its excellent performance in terms of accuracy and speed.

To track the faces of the subjects in the image plane, an extended Kalman filter (EKF) is employed~\cite{ribeiro2004kalman}. 
Define the \textit{state} vector of a target subject at time $k$, as \mbox{$\boldsymbol{x}_k = [x_k^c, y_k^c,\dot{x}_k^c, \dot{y}_k^c, h_k, d_k, \dot{d}_k]^T$}, where $x^c_k, y^c_k$ are the true coordinates of the center of his/her face in the thermal image, $\dot{x}_k^c, \dot{y}_k^c$ its velocities along the vertical and horizontal directions, $h_k$ is the true height of the bounding box enclosing the subject's face, $d_k$, the distance of the target from the camera in the physical space, and $\dot{d}_k$ its time derivative (rate of variation). 

The observation vector obtained from the YOLOv3 face detector, denoted by $\boldsymbol{z}_k = [\tilde{x}_k^c, \tilde{y}_k^c,\tilde{h}_k]^T$, contains noisy measurements of the face position and height (represented by the height of the bounding box), which are distinguished from their true values by the superscript \mbox{`` $\tilde{}$ ''}. Denote the observation noise by vector \mbox{$\boldsymbol{r}_{k}\sim \mathcal{N}\left(\boldsymbol{0}, \boldsymbol{R}\right)$}, with \mbox{$\boldsymbol{R} = \mathrm{diag}\left( \sigma^2_{\tilde{x}^c}, \sigma^2_{\tilde{y}^c}, \sigma^2_{\tilde{h}} \right)$}, with diagonal elements representing the (constant) observation noise variances of $\tilde{x}_k^c$, $\tilde{y}_k^c$ and $\tilde{h}_k$, respectively. In the implementation $\sigma^2_{\tilde{x}^c}=\sigma^2_{\tilde{y}^c}=0.01$ and $\sigma^2_{\tilde{h}}=20$ are used.

The EKF state transition model is defined as \mbox{$\boldsymbol{x}_{k+1} = f\left(\boldsymbol{x}_{k}, \boldsymbol{u}_{k}\right)$},
where $f(\cdot)$ is the transition function, connecting the system state at time $k$, $\boldsymbol{x}_k$, to that at time $k+1$, $\boldsymbol{x}_{k+1}$, and vector \mbox{$\boldsymbol{u}_k\sim \mathcal{N}\left(\boldsymbol{0}, \boldsymbol{Q}\right)$} represents the process noise. In the model used in this work, the process noise includes $4$ independent components, representing two random accelerations of the bounding-box center coordinates, $u_{k}^{x}, u_{k}^{y}$, a random noise term for the bounding-box dimension, $u_{k}^{h}$, and a random acceleration for the subject's distance, $u_{k}^{d}$. Therefore, it can be written $\boldsymbol{u}_k = \left[u_{k}^{x}, u_{k}^{y}, u_{k}^{h}, u_{k}^{d}\right]^T$ with covariance matrix $\boldsymbol{Q} = \mathrm{diag}\left[\sigma^2_{x}, \sigma^2_{y}, \sigma^2_{h}, \sigma^2_{d} \right]$. In the implementation, $\sigma^2_{x}=\sigma^2_{y}= \sigma^2_{d}=5$ and $\sigma_{h}=5.148$ are used (see \secref{sec:g-est}).

Assuming that the target moves according to a \textit{constant velocity} (CV) model, from the state definition it follows that 
\begin{equation} \label{eq:f-def}
	f\left(\boldsymbol{x}_{k}, \boldsymbol{u}_{k}\right) = 
	\left[
	\begin{array}{c}
		x_{k} + \Delta \dot{x}_{k} + u_{k}^{x}\Delta^2/2 \\
		y_{k} + \Delta \dot{y}_{k} + u_{k}^{y}\Delta^2/2 \\
		\dot{x}_{k} + u_{k}^{x}\Delta \\
		\dot{y}_{k} + u_{k}^{y}\Delta\\
		g\left(d_{k} + \Delta  \dot{d}_{k} + u_{k}^{d}\Delta^2/2\right) + u_{k}^{h}\\
		d_{k} + \Delta \dot{d}_{k} + u_{k}^{d}\Delta^2/2 \\
		\dot{d}_{k} + u_{k}^{d}\Delta\\
	\end{array}
	\right], 
\end{equation}
where the only non-linear term is function $g(\cdot)$, which relates the subject's distance extracted by the thermal camera to the height $h_k$ of the bounding-box enclosing his/her face. 
The proposed approach consists in \textit{(i)} obtaining an estimate for $g(\cdot)$ in an \textit{offline} fashion using training data, and \textit{(ii)} using such estimate in the EKF model. These two steps are detailed next.

\subsubsection{Estimation of function $g(\cdot)$}\label{sec:g-est}
Function $g(\cdot)$ maps the distance of the target from the thermal camera $d_k$, at time $k$, onto the corresponding height of the bounding box, $h_k$, as follows,
\begin{equation}
	h_k = g(d_k) + u^h_k.
\end{equation}
Using $N_t$ training samples $\{h_i, d_i\}_{i=1}^{N_t}$ containing the true distances of the target, $d_i$, and the measured bounding box height, $h_i$,  $g(\cdot)$ is obtained solving an \textit{offline} non-linear least-squares (LS) problem of the form
\begin{equation}\label{eq:ls-bb}
	\argmin_g \sum_{i=1}^{N_t}\left(h_i - g(d_i)\right)^2.
\end{equation}
From the equations of the pinhole camera model~\cite{princeCVMLI2012}, $g(\cdot)$ is restricted to the family of hyperbolic functions with shape \mbox{$g(d_i) = b_0 / (d_i+b_1) + b_2$}, reducing the problem to that of estimating the parameters $b_0$, $b_1$, and $b_2$, i.e.,
\begin{equation}\label{eq:ls-hyp}
	\argmin_{b_0, b_1, b_2} \sum_{i=1}^{N_t}\left(h_i - \frac{b_0}{d_i+b_1} + b_2\right)^2 .
\end{equation}
This optimization problem is here solved using the Levenberg-Marquardt algorithm~\cite{nocedal2006numerical} for non-linear LS fitting: with the experimental setup used in this paper, $b_0=162.04, b_1=0.61, b_2=-14.79$ are obtained.

Note that the process noise acts on the bounding-box dimension in two ways, inside the function $g(\cdot)$, modeling the uncertainty in the subject's distance due to the random acceleration, and through the additive term $u^h_k$, modeling the imperfect estimation of $g(\cdot)$ itself. The variance of $u^h_k$ can be estimated from the residuals, after fitting the training measurements with function $g(\cdot)$.

\subsubsection{Using $g(\cdot)$ in the EKF}\label{sec:g-use}

Due to the non-linear dependence of the state $\boldsymbol{x}_k$ on the process noise $\boldsymbol{u}_{k}$, in the EKF operations the following transformed process noise covariance matrix is used~\cite{simon2006optimal}
\begin{equation}\label{eq:q-prime}
	\boldsymbol{Q}'_k = \boldsymbol{L}_{k}\boldsymbol{Q}\boldsymbol{L}_{k}^T, \quad \mbox{with } \boldsymbol{L}_k = \frac{\partial f \left(\boldsymbol{x}_{k}, \boldsymbol{u}_{k}\right)}{\partial \boldsymbol{u}_{k}} \bigg \rvert_{\hat{\boldsymbol{x}}_{k|k}},
\end{equation}
where matrix $\boldsymbol{L}_k$ is the Jacobian of function $f(\cdot)$ with respect to the process noise vector, evaluated for the current state estimate. Using the above system model, the system state estimate at time $k$, $\hat{\boldsymbol{x}}_{k}$, is recursively obtained along with the corresponding error covariance matrix, $\boldsymbol{P}_{k}$.
By definition of the EKF state, this allows us to get a coarse estimate of the distance of the subjects from the TC, which is exploited in the radar-TC data association step, see \secref{sec:data-assoc}.

\subsection{Thermal Camera: Subject Temperature Estimation}
\label{sec:temp-est}

The body temperature is obtained from the thermal camera readings in the bounding-boxes contained in $\hat{\boldsymbol{x}}_{k}$, for each subject, and for all the time steps in which he/she is tracked by the EKF. At any given time $k$, a single (noisy) temperature measurement, $\tilde{T}_k$, is extracted by taking the maximum value across all the pixels in the current bounding box. Denoting by $B_k$ the $2$-D region of the image enclosed by the bounding box, and by $B_{ki}$ the intensity of its pixel $i$, it holds $\tilde{T}_k = \max_i B_{ki}$. 

\subsection{mmWave Radar: People Detection and Tracking}  
\label{sec:mmwave-track}

\begin{figure*}[t!]
	\begin{center}   
		\centering
		\subcaptionbox{ \label{fig:clust-1}}[4.2cm]{\includegraphics[width=4.2cm]{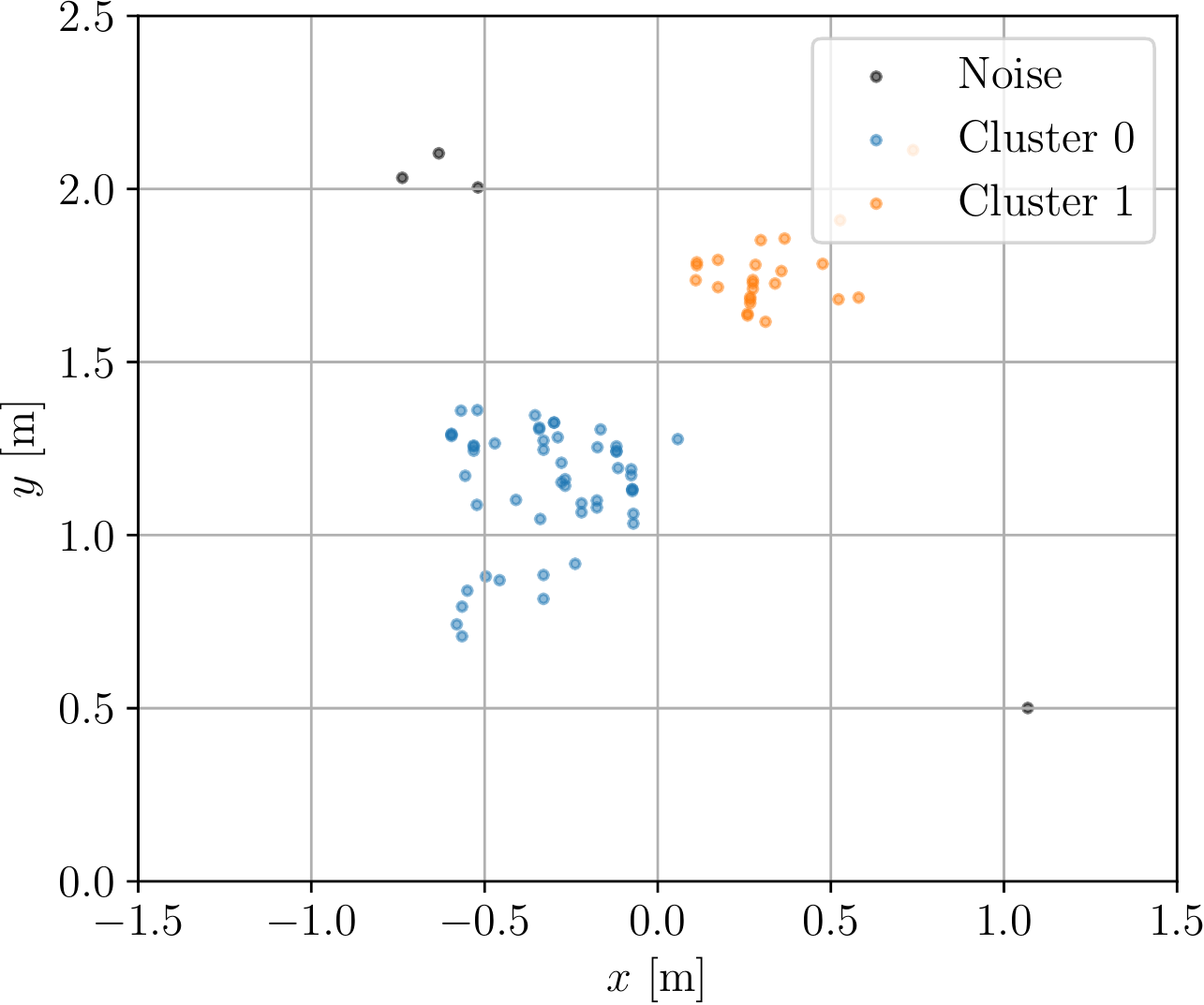}}
		\subcaptionbox{\label{fig:clust-2}}[4.2cm]{\includegraphics[width=4.2cm]{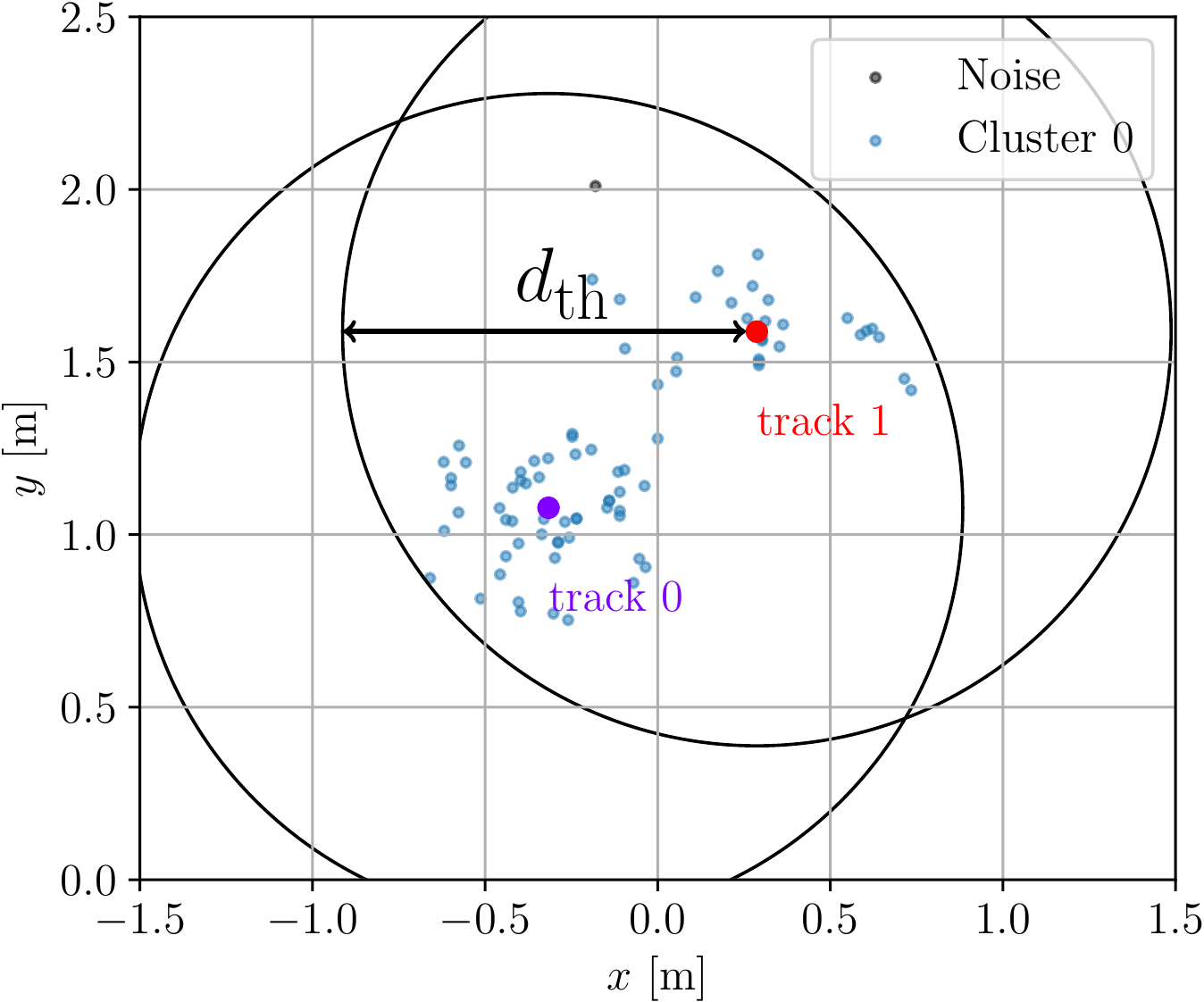}}
		\subcaptionbox{\label{fig:clust-3}}[4.2cm]{\includegraphics[width=4.2cm]{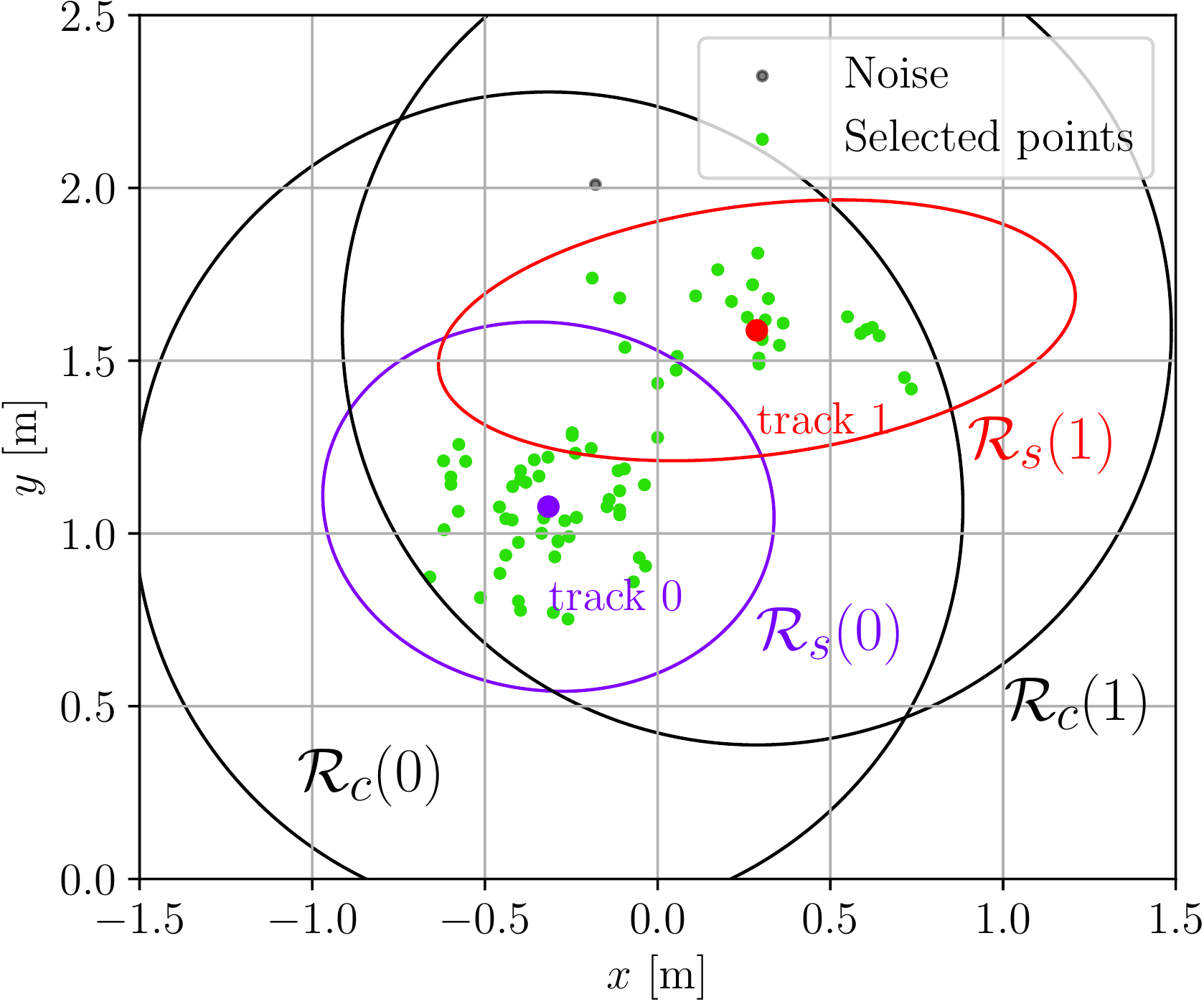}}
		\subcaptionbox{\label{fig:clust-4}}[4.2cm]{\includegraphics[width=4.2cm]{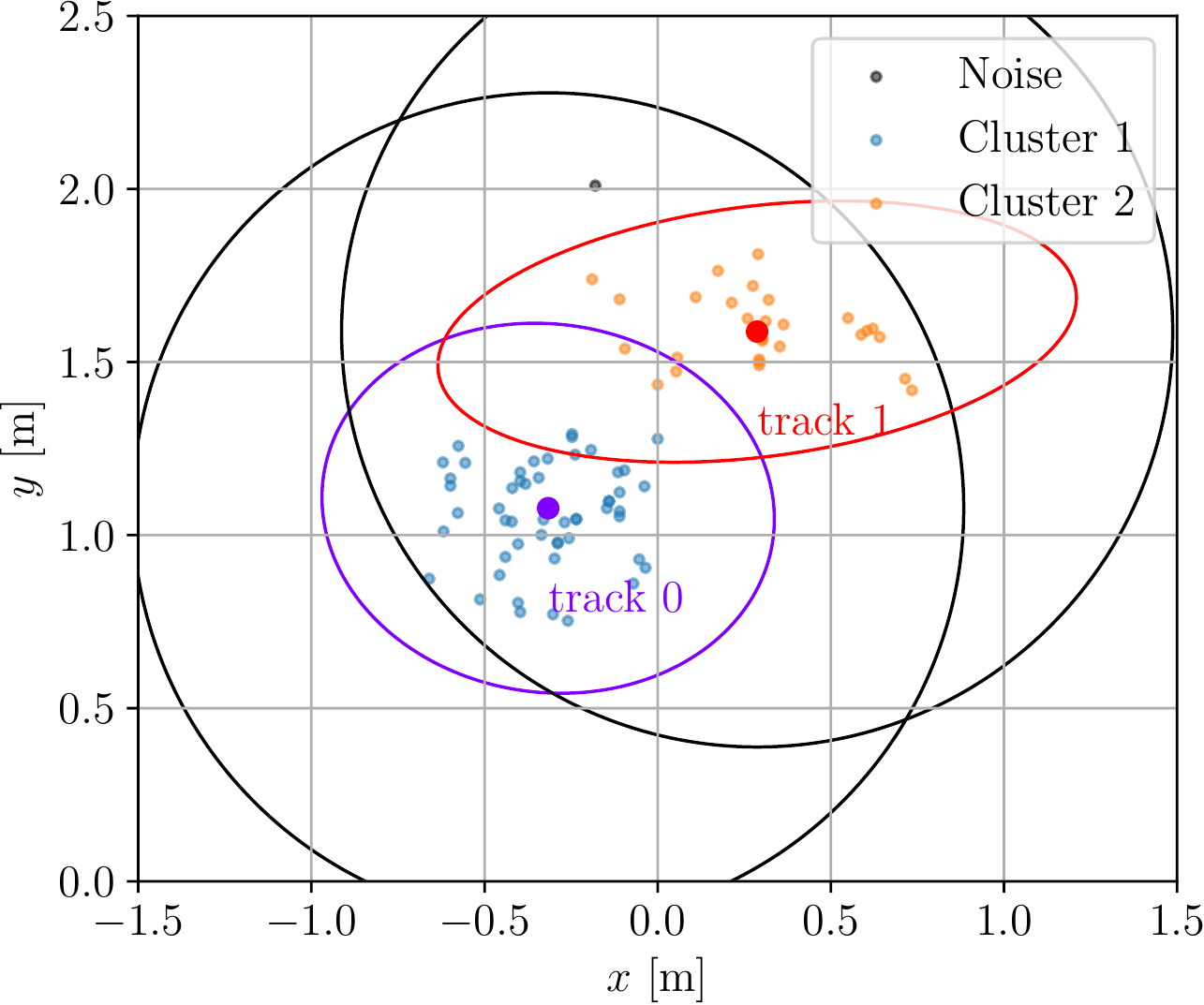}}
		\caption{Illustration of the proposed clustering method. In (a) the point-clouds belonging to $2$ subjects are well separated and DBSCAN outputs the correct clustering. In the next time-step, (b), DBSCAN fails and merges the two clusters into one. The proposed method selects the points to re-cluster using the tracks positions together with \eq{eq:region-c} and \eq{eq:region-s}, as shown in (c), and outputs the correct result using GM on the selected points with $n_{\mathcal{G}}=2$, see (d).}
		\label{fig:highly-acc-clust}
	\end{center}
	\vspace{-0.5cm}
\end{figure*}
The common approach to people tracking from mmWave radar point-clouds~\cite{zhao2019mid, meng2020gait, pegoraro2021realtime} includes 

\noindent\textit{(i)} \textbf{detection}: using density-based clustering to separate the points generated by the subjects from clutter and noise;

\noindent\textit{(ii)} \textbf{tracking}: applying Kalman filtering techniques~\cite{kalman1960new} on each cluster centroid to track the movement trajectory of each subject in space.

Detection is typically performed using \mbox{DBSCAN}~\cite{ester1996density}, an unsupervised density-based clustering algorithm that
takes two input parameters, $\varepsilon$ and $m_{\rm pts}$, respectively representing a radius around each point and the minimum number of other points that must be inside such radius to satisfy a certain density condition. Given the description of the radar measurements from \secref{sec:fmcw-radar}, the coordinates of the points in the horizontal plane ($x-y$) are used as input to DBSCAN, which outputs a list of detected clusters and a set of points which are classified as \textit{noise}. Typically, the centroid of each cluster is used as an \textit{observation} of the subject's position, feeding a subsequent KF tracking algorithm~\cite{zhao2019mid}.
DBSCAN has proven to be robust and accurate as long as the subjects do not come too close to one another~\cite{wagner2017radar, zhao2019mid, meng2020gait}, see also \fig{fig:clust-1}. When this occurs (\fig{fig:clust-2}), the algorithm often fails to distinguish between adjacent subjects, merging their contributions into a single cluster~\cite{feng2020evaluating}.


In the KF tracker, the \textit{state} of each subject at time $k$ is defined as $\boldsymbol{s}_k = \left[x_k, y_k, \dot{x}_k, \dot{y}_k\right]^T$, containing the $x-y$ subject's coordinates and the corresponding velocities.
The state evolution is assumed to obey $\boldsymbol{s}_k = \boldsymbol{A} \boldsymbol{s}_{k-1}$, where the transition matrix $\boldsymbol{A}$ represents a constant-velocity (CV) model~\cite{wagner2017radar}.
The KF computes an estimate of the state for a target subject at time $k$, denoted by $\hat{\boldsymbol{s}}_{k}$, by sequentially updating the predictions from the CV model with the new observations. The association between the new observations (time $k$) and the previous states (time $k-1$) exploits the nearest-neighbors joint probabilistic data association algorithm (NN-JPDA)~\cite{wagner2017radar, shalom2009probabilistic}.
milliTRACE-IR also uses the just described DBSCAN and KF based signal processing pipeline, but improves it significantly with a novel clustering procedure to better resolve the point-clouds of subjects that are close to one another. 
The designed solution to enhance the tracking accuracy in such cases is a major contribution of the present work and is detailed next. 

\subsection{mmWave Radar: Highly Accurate Clustering}
\label{sec:cluster-res}

As a possible solution to DBSCAN drawbacks, one may adjust the parameters $\varepsilon$ and $m_{\rm pts}$ so as  to correctly resolve the clustering ambiguity, even for closely spaced targets. However, $\varepsilon$ and $m_{\rm pts}$ interact in a complex and often unpredictable way, making the design of such adaptation rule difficult.

milliTRACE-IR adopts a different approach, which combines \textit{(i)} the standard DBSCAN algorithm with fixed $\varepsilon$ and $m_{\rm pts}$, \textit{(ii)} the spatial locations of the subjects, available from the tracking procedure, and \textit{(iii)} the Gaussian mixture (GM) clustering algorithm~\cite{bishop2006pattern}. 
The designed algorithm, reported in \alg{alg:prop-cluster} and exemplified in \fig{fig:highly-acc-clust}, proceeds as follows. At first, the DBSCAN algorithm is applied to obtain an estimate of the clusters and a reasonable separation between the noise points and those belonging to actual subjects, using $\varepsilon=0.4$~m and $m_{\rm pts}=10$. DBSCAN outputs a cluster label for each point $\boldsymbol{p} \in \mathcal{P}_k$, denoted by $\ell_{\boldsymbol{p}}$. Clusters are denoted by $\mathcal{C}_n$, and their centroids by $\bar{\boldsymbol{c}}_n$, with $n=1, \dots, n_k$.

The next step is to identify which of the tracked subjects get closer than a critical distance $d_{\rm th}$ from one another. The clusters provided by standard DBSCAN for these subjects are expected to be incorrect, as the point-cloud data from these would be merged into a single cluster. To pinpoint these subjects, their KF state is leveraged, which corresponds to a filtered representation of their trajectories. Consider track $t$ at time $k$, its coordinates are predicted as \mbox{$\hat{\boldsymbol{s}}_{k}^t = \boldsymbol{A}\hat{\boldsymbol{s}}_{k-1}^t$} (see line $2-3$ in \alg{alg:prop-cluster}). For any two subjects with associated tracks $t$ and $t^\prime$, milliTRACE-IR checks whether $||\hat{\boldsymbol{s}}_{k}^t - \hat{\boldsymbol{s}}_{k}^{t^\prime}||_2 < d_{\rm th}$.
If this occurs, as shown in the example of \fig{fig:clust-2} for tracks $t=0$ and $t^\prime=1$, $t$ and $t^\prime$ are termed {\it nearby subjects}. Hence, define $\mathcal{G}$ as the set of subjects that are mutually within a radius of $d_{\rm th}$ from one another. A group $\mathcal{G}$ can be constructed starting from any subject and recursively adding all the subjects who are closer than $d_{\rm th}$ from any of the set members. If a subject has no other subjects within distance $d_{\rm th}$, it will be the only member of his group. Collecting all the disjoint groups, constructed from the maintained tracks at time $k$, set $\boldsymbol{\mathcal{G}}_k(d_{\rm th})$ is obtained, containing all the nearby subjects groups. Once the nearby groups are identified, the ambiguities inside each group $\mathcal{G}$ containing more than one member are resolved by recomputing the clustering labels as follows. Consider a single group $\mathcal{G}$. 
\begin{algorithm}[t!]
	\caption{Clustering refinement method.}
	\label{alg:prop-cluster}
	\begin{algorithmic}[1]
		\REQUIRE States of the targets at time $k-1$, observed point-cloud at time $k$, $\mathcal{P}_k$.
		\ENSURE Labels $\ell_{\boldsymbol{p}}$, $\forall \, \boldsymbol{p} \in \mathcal{P}_k$.
		\STATE $\left\{\ell_{\boldsymbol{p}} \right\}_{\boldsymbol{p} \in \mathcal{P}_k}, \left\{\mathcal{C}_n\right\}_{n=1}^{n_k}\leftarrow \mathrm{DBSCAN}(\varepsilon, m_{\rm pts}, \mathcal{P}_k)$
		\STATE \mbox{$\hat{\boldsymbol{s}}_k^t \leftarrow \boldsymbol{A}\hat{\boldsymbol{s}}_{k-1}^t$} all maintained tracks $t$
		\STATE Find groups of nearby subjects $\boldsymbol{\mathcal{G}}_k (d_{\rm th})$ 
		\FOR{each $\mathcal{G} \in \boldsymbol{\mathcal{G}}_k(d_{\rm th})$}
		\STATE $n_{\mathcal{G}} \leftarrow |\mathcal{G}|$
		\IF{$n_{\mathcal{G}} > 1$}
		\STATE $\mathcal{R}(\mathcal{G}) \leftarrow \bigcup_{t \in \mathcal{G}} (\mathcal{R}_{c}(t) \cap \mathcal{R}_{s}(t))$
		\STATE $\mathcal{S} \leftarrow \left\{\boldsymbol{p} \in \mathcal{C}_n \mbox{ such that } \bar{\boldsymbol{c}}_n \in \mathcal{R}(\mathcal{G})\right\}$
		\STATE discard $\ell_{\boldsymbol{p}}, \forall \, \boldsymbol{p} \in \mathcal{S}$
		\STATE $\left\{\ell_{\boldsymbol{p}}\right\}_{\boldsymbol{p} \in \mathcal{S}}, \left\{\pi_q \right\}_{q=1}^{n_{\mathcal{G}}}\leftarrow \mathrm{GM}(n_{\mathcal{G}}, \mathcal{S})$
		\STATE discard cluster $q$ if $\pi_q < \pi_{\rm thr}$
		\ENDIF
		\ENDFOR
	\end{algorithmic}
\end{algorithm}
To delimit the region where the clustering has to be refined, the following additional regions are defined. The sample covariance matrix of the \textit{last} cluster associated with track $t$ is denoted by $\boldsymbol{\Sigma}_n^t$, and contains information about the shape of the subject's cluster. The regions of the plane containing the points that are within a radius of $d_{\rm th}$ from $\hat{\boldsymbol{s}}_{k}^t$, can be written as
\begin{equation}\label{eq:region-c}
	\mathcal{R}_{c}(t) =  \left\{\boldsymbol{x} \in \mathbb{R}^2
	\mbox{ s.t. }
	\left|\left|\boldsymbol{x} - \hat{\boldsymbol{s}}_{k}^t\right|\right|_2 < d_{\rm th}\right\},
\end{equation}
and the regions of points with a squared Mahalanobis distance smaller than $\gamma$ are
\begin{equation}\label{eq:region-s}
	\mathcal{R}_{s}(t) =  \left\{\boldsymbol{x} \in \mathbb{R}^2 \mbox{ s.t. }
	\left|\left|\boldsymbol{x} - \hat{\boldsymbol{s}}_{k}^t\right|\right|_{\left(\boldsymbol{\Sigma}_n^t\right)^{-1}}^2< \gamma \right\}.
\end{equation}
In the implementation, $d_{\rm th} = 1.2$~m and $\gamma = 9.21$ were used.\footnote{The value of $\gamma$ corresponds to a probability of $99\%$ of falling inside the region, assuming that the points in the cluster are distributed on the plane according to a Gaussian distribution around $\hat{\boldsymbol{s}}_{k}^t$.} 
Then, the labels assigned by DBSCAN to all the points belonging to a cluster whose centroid falls inside region $\mathcal{R}(\mathcal{G}) = \cup_{t \in \mathcal{G}} ( \mathcal{R}_{c}(t) \cap \mathcal{R}_{s}(t))$, are discarded (lines $7-9$ in \alg{alg:prop-cluster}).\footnote{Discarding a label corresponds to setting it equal to that used by DBSCAN to represent noise points.} This set of points is denoted by $\mathcal{S}$.

Then, the GM algorithm is applied to the points belonging to set $\mathcal{S}$ to refine the clusters within this region, see the green points in \fig{fig:clust-3}. As GM requires the number of clusters to be specified in advance, it is set to be equal to the number of subjects in the group, i.e., $n_{\mathcal{G}} = |\mathcal{G}|$. The GM algorithm outputs the labels $\ell_{\boldsymbol{p}}$ for each point $\boldsymbol{p} \in \mathcal{S}$ and the weight of the Gaussian component associated with each GM cluster, $\pi_q \in [0,1], q=1, \dots, n_{\mathcal{G}}$, with $\sum_q \pi_q = 1$. The new labels are used to replace the ones previously found by DBSCAN (\fig{fig:clust-4}), unless the GM clusters have very small weights, i.e., the new clusters having $\pi_q < \pi_{\rm thr}$ are discarded and treated as noise points. The threshold value used in the implementation is $\pi_{\rm thr} = 0.1 / n_{\mathcal{G}}$.

The proposed method effectively solves the problem faced by DBSCAN in resolving subjects close to one another. The cost of this improvement is that an additional GM algorithm has to be applied to a subset of the point-cloud, however, at each time $k$ the number of points in this subset is typically much smaller than that in the full point-cloud $\mathcal{P}_k$.

\subsection{Radar and Thermal Camera Data Association}\label{sec:data-assoc}

Upon tracking the subjects in the TC image plane and in the physical space, respectively using the measurements from the TC and from the mmWave radar sensor, a track-to-track association method is applied to link the movement trajectory of each person to his/her body temperature.

Assume that, at time $k$, the system has access to $N_k^{\rm rad}$ tracks from the radar sensor and $N_k^{\rm tc}$ tracks from the thermal camera, indicized by $i$ and $j$, respectively. The data association strategy used in milliTRACE-IR consists in \textit{(i)} computing a \textit{cost} for each association \mbox{$(i \leftrightarrow  j)$}, and \textit{(ii)} solving the resulting combinatorial cost minimization problem to associate the best matching track pairs.
The main challenge in the association of radar and thermal camera tracks is the design of a cost function that grants robustness in the presence of multiple targets, which may enter the monitored area in unpredictable ways.
The key point is to gauge the similarity of the tracks by comparing them in terms of common quantities, which can be estimated from both devices.

\begin{figure}[t!]
\begin{center}   
\includegraphics[width=8.6cm]{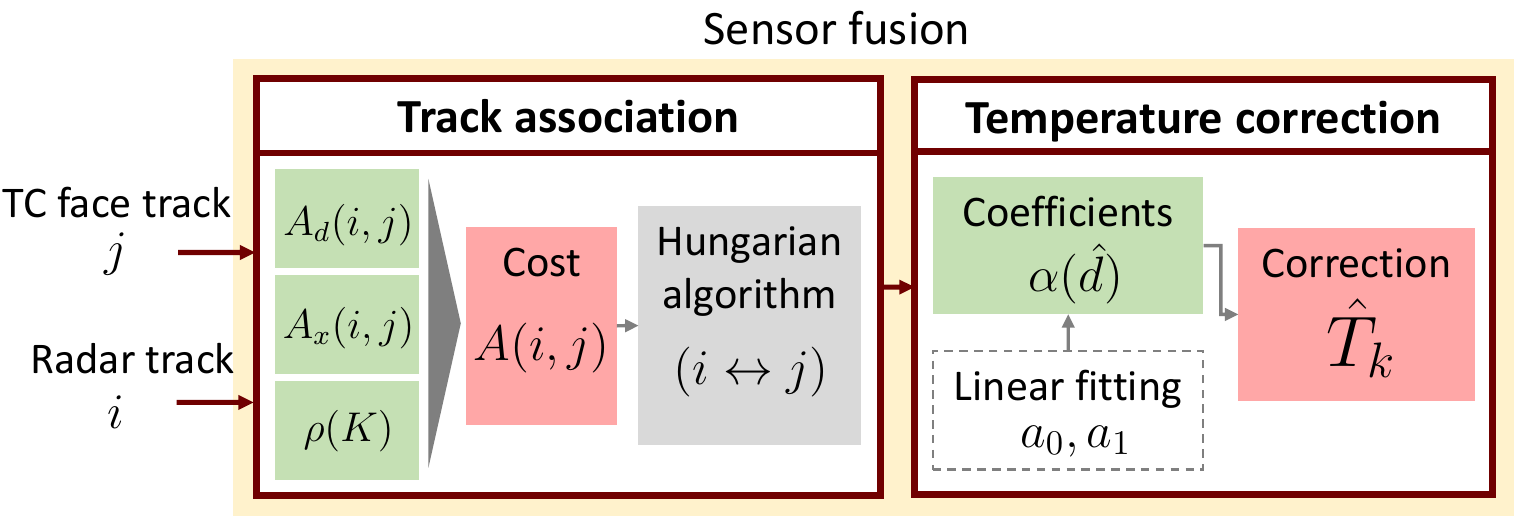} 
\caption{Block diagram of the sensor fusion step.}
\label{fig:sensor-fusion}
\end{center}
\end{figure}

Assume also that the two sensors are located in the same position and with the same orientation (co-located). In this setup, \textit{(i)} the \emph{distance} between the subjects and the sensors is the same, so its estimate should match for tracks representing the same subjects, and \textit{(ii)} the radar KF states containing the coordinates of the subjects' positions can be projected onto the TC image plane; after this operation, the horizontal component of the radar projections and the horizontal component of the TC bounding boxes position should match for correctly associated tracks.
\begin{figure}[t]
	\centering
	\subcaptionbox{ \label{fig:Ad_example}}[3.8cm]{\includegraphics[width=3.5cm]{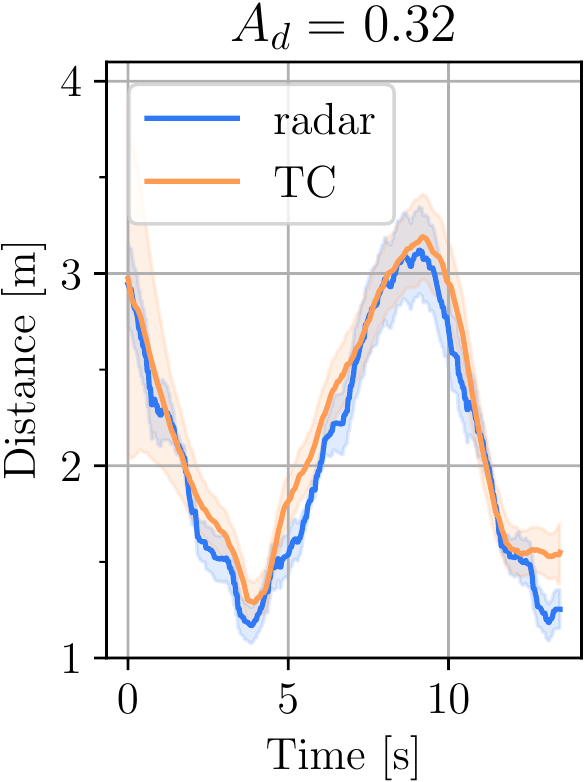}}
	\subcaptionbox{ \label{fig:Ax-example}}[4.1cm]{\includegraphics[width=3.8cm]{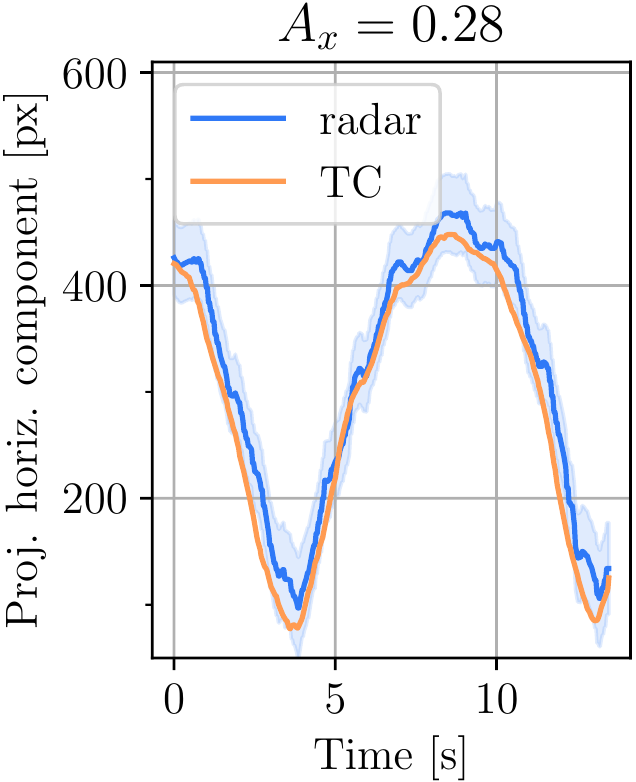}}
	\caption{Example of distance (a) and horizontal projection (b) estimates from a track. The shaded areas represent the standard deviations. The corresponding values for $A_d$ and $A_x$ are shown above.}
	\label{fig:assoc-example}
\end{figure}
To reliably associate radard and TC tracks, milliTRACE-IR uses a cost function consisting of the following components, see also \fig{fig:sensor-fusion}\\
\noindent \textbf{Estimated distance cost.} Denote by $d_k^i$ the estimated distance of radar track $i$, and by $d_k^j$ the estimated distance of TC track $j$. 
Recalling that $\hat{\boldsymbol{s}}_{k}^i$ is the position of subject $i$ at time $k$, $d_k^i$ is computed using Pythagora's formula as \mbox{$d_k^i = \sqrt{\left(\hat{\boldsymbol{s}}_{k}^i\right)_1^2 + \left(\hat{\boldsymbol{s}}_{k}^i\right)_2^2}$}. Distance $d_k^j$, instead, is retrieved directly from the tracking state of the TC. Considering $K$ subsequent time steps where radar track $i$ and TC track $j$ are both available, the estimated distance cost is defined as 
\begin{equation}\label{eq:dist-cost}
A_d(i, j) = \frac{1}{K} \sum_{k=1}^{K} \frac{(d^i_k - d^j_k)^2}{\sigma^2_{d^i_k} + \sigma^2_{d^j_k}},  \quad \enspace
\end{equation}
where $\sigma^2_{d^i_k}$ and $\sigma^2_{d^j_k}$ represent the variances of the two distance estimates. An illustrative example is shown in \fig{fig:Ad_example}.\\
\noindent \textbf{Projected horizontal component cost.} The horizontal components of the radar state projection and of the TC bounding box center are respectively denoted by $x_k^i$ and $x_k^j$. The radar positions provided by the KF state have only two dimensions, $x$ and $y$ (the first and second components of the state vector). However, \mbox{three-dimensional} vectors are needed for their proper projection onto the TC image plane. For this reason, a $0$-valued $z$ component is artificially added, under the assumption that the subjects' position at height $0$ is the one being tracked. For this, an augmented subject's position vector, $\boldsymbol{a}_k^i = [\left(\hat{\boldsymbol{s}}_{k}^i\right)_1, \left(\hat{\boldsymbol{s}}_{k}^i\right)_2, 0]^T$, is defined. $x_k^i$ is computed by projecting the radar coordinates $\boldsymbol{a}_k^i$ onto the TC image plane, as $\boldsymbol{a}_k^{i, {\rm proj}} = \boldsymbol{\Psi} \boldsymbol{a}_k^i$ (see \secref{sec:TC_preliminaries}), applying to it a radial distortion based on the estimated distortion coefficients and retaining only the \mbox{$x$-axis} component. Projection $x_k^j$ corresponds to the $x$ coordinate of the TC tracked state.
The projected horizontal component cost is defined, for $K$ subsequent time steps of radar track $i$ and TC track $j$, as
\begin{equation}\label{eq:xproj-cost}
A_x(i, j) = \frac{1}{K}\sum_{k=1}^{K} \frac{(x_k^i - x_k^j)^2}{\sigma^2_{x_k^i} + \sigma^2_{x_k^j}},  \quad \enspace
\end{equation}
where $\sigma^2_{x^i_k}$ and $\sigma^2_{x^j_k}$ are the variances of the two estimates.
An illustrative example is shown in \fig{fig:Ax-example}.\\
\noindent \textbf{Track length coefficient.} Recalling that $\Delta$ is the (constant) sampling interval, the proposed cost function accounts for the length $K$ of the tracks that are to be associated, favoring longer tracks. To this aim, the following coefficient is defined,
\begin{equation}
\label{eq:rho}
\rho(K) = \frac{1}{\log(K\Delta)}.
\end{equation}
Note that $\rho(K)$ is a weight factor for a cost (see the later \eq{eqn:cost-function}), which decreases with the track length $K$. This means that a smaller cost is implied when the associated tracks $i$ and $j$ are longer. Also, in the implementation, it holds $K > 1/\Delta$, so $\rho(K)$ is always positive. \\
\noindent \textbf{Association cost function for radar and TC tracks.} The association cost $A(i,j)$ for the tracks pair $(i, j)$ ($i$ refers to a radar track and $j$ to a TC track) is obtained summing \eq{eq:dist-cost} and \eq{eq:xproj-cost}, to gauge how well the two tracks match in terms of their estimated distance across time, and estimated position on the horizontal projected axis on the TC image plane, respectively. The sum is then weighted by the coefficient of \eq{eq:rho}. Formally, $A(i,j)$ is given by
\begin{equation}
\label{eqn:cost-function}
A(i, j)  =  \rho(K) \left[ A_d(i, j) + A_x(i, j)\right].
\end{equation}
Costs $A(i,j)$, $i=1,\,\dots,\,N_k^{\rm rad}$, $j=1,\,\dots,\,N_k^{\rm tc}$, are arranged into an \mbox{$N_k^{\rm rad} \times N_k^{\rm tc}$} matrix, and the optimal association of tracks is obtained by minimizing the overall cost, computed through the Hungarian algorithm~\cite{kuhn1955hungarian}. The Hungarian algorithm takes the cost matrix as input and solves the problem of pairing each radar track with a single TC track (by minimizing the total cost), with an overall complexity of $O((N_k^{\rm rad}N_k^{\rm tc})^3)$.\\
In general, the radar and the TC would be deployed at different spatial locations. However, knowing their relative position and orientation, a roto-translation matrix $\boldsymbol{\Phi}$ can be obtained to geometrically transform the data into a new coordinate system where the TC and the radar sensors are co-located, as described above. In this work, the TC position and orientation are selected as the reference coordinate system, and the positions estimated from the radar sensor are transformed into it.

\subsection{Temperature Correction}\label{sec:temp-corr}
In line with~\cite{savazzi2020processing}, the direct reading of each subject's temperature , $\tilde{T}_k$, is subject to a scaling factor, $\alpha(d_k)$, with respect to the true temperature $T$, where $\alpha(d_k)$ depends on the distance from the TC, i.e.,
\begin{equation}
T = \alpha(d_k)\tilde{T}_k .
\end{equation}
For an accurate temperature screening, the scaling factor $\alpha(d_k)$ is estimated from the training data, considering a linear model of the form
\begin{equation}
\label{eq:alpha_fit}
\alpha(d_k) = a_0 + a_1 d_k.
\end{equation}
Using $N_t^'$ training measurements $\{\tilde{T}_i, d_i, T\}_{i=1}^{N_t^'}$, the fitting coefficients $a_0, a_1$ are obtained by solving 
\begin{equation}\label{eq:alpha-params}
\argmin_{a_0, a_1} \sum_{i=1}^{N_t^'}\left(T - \alpha(d_i)\tilde{T}_i\right)^2.
\end{equation}
From the above optimization problem, in this work the above parameters are set to \mbox{$a_0=1.116$}, \mbox{$a_1=0.013$}. At system operation time, denoting by $M$ the number of time-steps for which the subject is correctly tracked by the EKF, his/her true temperature at time $k$ is finally estimated as
\begin{equation}
\label{eq:corrected-temp}
\hat{T}_k = \frac{1}{M}\sum_{j=k-M+1}^{k} \alpha(\hat{d}_j)\tilde{T}_j,
\end{equation}
where $\alpha(\cdot)$ is defined in \eq{eq:alpha_fit}, using the parameters obtained from \eq{eq:alpha-params}, while
$\hat{d}_j$ is an estimate of the distance obtained by the system at time-step $j$. To improve the temperature estimates, milliTRACE-IR performs sensor fusion by exploiting the association between the TC face tracks and the mmWave radar tracks (see \secref{sec:data-assoc}). In \eq{eq:corrected-temp}, the coefficients $\alpha(\hat{d}_j)$ are computed using the distances estimated by the mmWave radar device, as these are much more accurate than those obtained from the TC. The impact of combining the temperature information from the TC and the accurate distance estimation capabilities of the radar is investigated in \secref{sec:temp-screen}. The block diagram for the temperature correction step is shown in \fig{fig:sensor-fusion}.

\subsection{Extraction of Feature Vectors from mmWave Point-Clouds}
\label{sec:dl-fextr}

\begin{table}[t!] 
\footnotesize
\caption{Summary of the architecture and training parameters of the NN used for gait feature extraction. \label{tab:nn-details}}
\begin{center}
\begin{tabular}{lccc}
	\toprule
	\multicolumn{4}{c}{{\bf Architecture }} \\
	\midrule
	\multicolumn{2}{c}{{\bf Layer/block }} & \multicolumn{2}{c}{{\bf Size }}  \\
	\cmidrule(lr){1-2}\cmidrule(lr){3-4}
	\multicolumn{2}{l}{PC features \cite{pegoraro2021realtime}} & \multicolumn{2}{c}{$3+2$ shared MLPs, ($98$, $196$)}\\
	\multicolumn{2}{l}{Temporal conv. \cite{pegoraro2021realtime}} & \multicolumn{2}{c}{$3$ Conv. $(3\times3)$, $32, 64, 128$ filt. }\\
	\multicolumn{2}{l}{Temporal conv. \cite{pegoraro2021realtime}} & \multicolumn{2}{c}{$3$ Conv. $(3\times3)$, $256, 128, 32$ filt. }\\
	\multicolumn{2}{l}{Global average pooling} & \multicolumn{2}{c}{$32$}\\
	\multicolumn{2}{l}{Fully connected} & \multicolumn{2}{c}{$32$}\\
	\multicolumn{2}{l}{$L_2$~normalization} & \multicolumn{2}{c}{$32$}\\
	\multicolumn{2}{l}{Fully connected} & \multicolumn{2}{c}{$16$}\\
	\midrule 
	\multicolumn{4}{c}{{\bf Training parameters }} \\
	\midrule
	Learning rate& && $10^{-4}$\\
	Optimizer&&& Adam \cite{goodfellow2016deep}\\
	Number of epochs &&&  $250$\\
	$\mathcal{L}_{\rm cen}$ weight, $\omega$&&&  $0.5$\\
	$L_2$-regularization parameter &&& $8\times10^{-5}$\\
	Dropout rate &&& $0.4$ \\
	Triplet margin, $\mu$ &&& $1$\\
	\bottomrule
\end{tabular} 		
\end{center}
\end{table}

\begin{figure*}[t!]
\begin{center}   
\includegraphics[width=17.5cm]{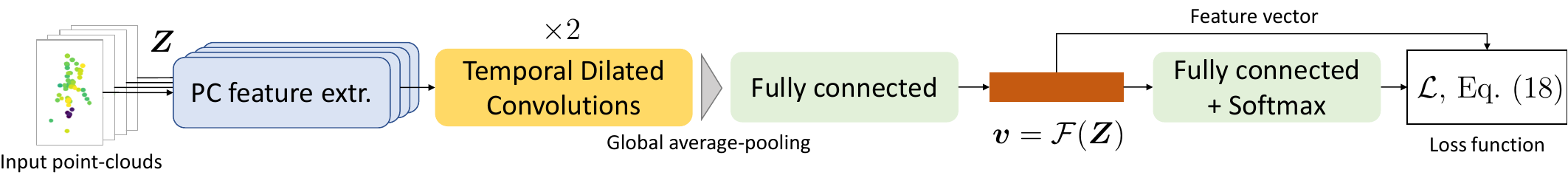} 
\caption{Block diagram of the NN feature extractor.}
\label{fig:nn}
\end{center}
\end{figure*}

To extract the gait features of the subjects, the neural network (NN) proposed in~\cite{pegoraro2021realtime}, which was originally developed for person identification, is here adapted. The network uses a point-cloud feature extraction block inspired by PointNet~\cite{qi2017pointnet}, and followed by temporal dilated convolutions~\cite{oord2016wavenet} to capture features related to the movement evolution in time. The proposed NN takes as input a radar point-cloud sequence, denoted by $\boldsymbol{Z}$, and outputs the corresponding feature vector $\boldsymbol{v} = \mathcal{F}(\boldsymbol{Z})$. \fig{fig:nn} shows the block diagram of the NN. First, the network is expanded with respect to~\cite{pegoraro2021realtime}, using augmented point-cloud feature extraction blocks composed of $3$ shared multi-layer perceptrons (MLPs) of size $98$ and $2$ MLPs of size $196$, yielding point cloud features of size $196 \times 1$. Then, $2$ temporal convolution blocks are used, containing $3$, $3\times 3$, convolutional layers each, with $(32,64,128)$ and $(256,128,32)$ filters, respectively, for the two blocks, and dilation rates of $1,2,4$ for the $3$ layers in each block. Then, after applying the same global average pooling operation of~\cite{pegoraro2021realtime}, a fully connected layer~\cite{goodfellow2016deep} is introduced before the classification output, which produces a vector $\tilde{\boldsymbol{v}}$ of dimension $32$.
The final feature vector is obtained using $L_2$-normalization on $\tilde{\boldsymbol{v}}$, i.e., $\boldsymbol{v} = \tilde{\boldsymbol{v}} / ||\tilde{\boldsymbol{v}}||_2$. A summary of the NN layers and their parameters is provided in \tab{tab:nn-details}.

\subsubsection{Training} the NN is trained to produce representative feature vectors, $\boldsymbol{v}$, containing information on the way of walking of the subjects. This requires that the network generalizes well to subjects \textit{not seen} at training time, as the performance of the re-identification mechanism strongly depends on the quality of the extracted features. To this end, in this work the NN is trained using a weighted combination of the \textit{cross-entropy loss}~\cite{goodfellow2016deep}, denoted by $\mathcal{L}_{\rm ce}$, the \textit{center loss}~\cite{wen2016discriminative}, $\mathcal{L}_{\rm cnt}$, and the \textit{triplet loss}~\cite{schroff2015facenet}, $\mathcal{L}_{\rm tri}$. 

The cross-entropy is the most widely used loss for classification purposes in deep learning, and here it is used to train the network to distinguish among the different subjects~\cite{goodfellow2016deep}. However, just training the NN on a classification problem does not lead to sufficiently discriminative features for the re-identification mechanism. The center loss is adopted to additionally force the feature representations belonging to the same class to be close in the feature space, in terms of Euclidean distance. Specifically, denoting by $\boldsymbol{c}_l$ the centroid of the feature vectors belonging to class $l$, the center loss is
\begin{equation}\label{eq:center}
\mathcal{L}_{\rm cen}(\boldsymbol{v}, l) = ||\boldsymbol{v} - \boldsymbol{c}_{l}||^2_2,
\end{equation}
where the centroids are learned as part of the training process via the back-propagation algorithm~\cite{wen2016discriminative}.

The triplet loss is used to push apart the feature representations of inputs belonging to different classes. For this, triplets of input samples are selected from the training set, two of them from the same class, leading to feature vectors $\boldsymbol{v}_a$ and $\boldsymbol{v}_b$, and one belonging to a different class, leading to a third feature vector $\boldsymbol{v}_c$. For further details on the triplet selection process, see Section~$3.2$ of~\cite{schroff2015facenet}. The triplet loss is written as
\begin{equation}\label{eq:triplet}
\mathcal{L}_{\rm tri}(\boldsymbol{v}_a, \boldsymbol{v}_b, \boldsymbol{v}_c) = \max \left\{||\boldsymbol{v}_a - \boldsymbol{v}_b||_2^2 - ||\boldsymbol{v}_a - \boldsymbol{v}_c||_2^2 + \mu, 0\right\},
\end{equation}
where $\mu$ is a margin hyperparameter, set to $1$. Hence, the feature extractor is trained with the following total loss function 
\begin{equation}\label{eq:loss}
\mathcal{L} = \mathcal{L}_{\rm ce} + \mathcal{L}_{\rm tri} + \omega \mathcal{L}_{\rm cen},
\end{equation}
where the parameter $\omega=0.5$ weighs the relative importance of the center loss. In the implementation, a training dataset containing mmWave radar point-clouds from $16$ subjects is used. It was collected in different indoor environments to increase the generalization capabilities of the NN. The optimization is carried out using Adam~\cite{goodfellow2016deep} with learning rate $10^{-4}$ and an $L_2$~regularization rate of $8\times 10^{-5}$ for $250$ epochs, as summarized in \tab{tab:nn-details}. Hyperparameters tuning was carried out using a greedy search procedure, optimizing the value of the loss $\mathcal{L}$ on a validation set containing a randomly selected subset ($20$\%) of the training data.

\subsubsection{Feature extraction} at inference time, i.e., during the system operation, the NN is used to compute feature vectors that are representative of the subjects' gait. Specifically, $45$ steps ($3$~seconds) long sequences of radar point-clouds are collected for each tracked subject. The point-cloud sequences are denote by $\boldsymbol{Z}$ in the following. The inner representation $\boldsymbol{v} = \mathcal{F}(\boldsymbol{Z})$, after \mbox{$L_2$-normalization}, is used as the feature vector for the following re-identification mechanism. 

\subsection{Weighted Extreme Learning Machine (WELM)}
\label{sec:elm-prel}

The weighted extreme learning machine (WELM)~\cite{zong2013weighted} is a particular kind of single-layer feedforward neural network in which the weights of the hidden nodes are chosen randomly, while the parameters of the output layer are computed analytically. Consider an $n_{\rm cls}$-class classification problem, a training set $\mathcal{V} = \cup_{n=1}^{n_{\rm cls}} \mathcal{V}_n$ of input {\it feature vectors} $\boldsymbol{v}$ (see \secref{sec:dl-fextr}), each with an associated one-hot encoded label $\boldsymbol{y}\in \{0, 1\}^{n_{\rm cls}}$, where $\mathcal{V}_n$ is the set containing the vectors from class $n=1,\dots,n_{\rm cls}$. For any $\boldsymbol{v} \in \mathcal{V}$, the WELM computes the matrix of hidden feature vectors $\boldsymbol{H} \in \mathbb{R}^{|\mathcal{V}| \times L}$, with rows $\boldsymbol{h}(\boldsymbol{v})$,
where $L$ is the number of WELM hidden units and $\boldsymbol{h}(\cdot)$ is a non-linear activation function. milliTRACE-IR uses $\boldsymbol{h}(\boldsymbol{v}) = \mathrm{ReLU}(\boldsymbol{W}\boldsymbol{v} + \boldsymbol{b})$ where $\mathrm{ReLU}$ is the rectified linear unit~\cite{goodfellow2016deep} ($\mathrm{ReLU}(x) = \max(x,0)$) and $\boldsymbol{W}, \boldsymbol{b}$ are the weights and biases of the ELM hidden layer, respectively. The elements of $\boldsymbol{W}$ and $\boldsymbol{b}$ are here generated from $\mathcal{N}(0, 0.1)$.
The WELM learning process amounts to computing, for each class $n$, the optimal values of an output weight vector $\boldsymbol{\beta}_n$ that minimizes the \textit{weighted} LS \mbox{$L_2$-regularized} quadratic cost function $||\boldsymbol{H}\boldsymbol{\beta}_n - y_n||_{\boldsymbol{\Omega}}^2 + \lambda ||\boldsymbol{\beta}_n||_2^2$, where $\lambda$ is a regularization parameter and $\boldsymbol{\Omega}$ is a diagonal weighting matrix used to boost the importance of those samples belonging to under-represented classes. This compensates for the tendency of the standard ELM to favor over-represented classes at inference time~\cite{zong2013weighted}. In the analyzed scenario, the individuals move freely in the environment across different rooms, so the number of feature vectors collected from each of them is not only unknown in advance, but highly variable. Hence, the training set usually contains unbalanced classes, and milliTRACE-IR uses
\begin{equation}\label{eq:weight-matrix}
\Omega_{i, i} = 1/|\mathcal{V}_{n_i}|, \, i = 1, \dots, |\mathcal{V}|,
\end{equation}
where $n_i = \argmax_n (\boldsymbol{y}_i)_n$ denotes the class of the $i$-th vector.
Stacking all the $\boldsymbol{\beta}_n$ into a single matrix $\boldsymbol{B} \in \mathbb{R}^{L\times n_{\rm cls}}$ and the labels into matrix $\boldsymbol{Y}\in \{0, 1\}^{|\mathcal{V}|\times n_{\rm cls}}$, the WELM output weights $\boldsymbol{B}$ can be computed in closed-form using one of the following equivalent expressions
\begin{align}
\boldsymbol{B} & = \boldsymbol{H}^T \left(\lambda \boldsymbol{I} + \boldsymbol{\Omega}\boldsymbol{H}\boldsymbol{H}^T\right)^{-1} \boldsymbol{\Omega}\boldsymbol{Y}, \label{eq:wbeta-nsmaller} \mbox{ or}\\
\boldsymbol{B} &= \left(\lambda \boldsymbol{I} + \boldsymbol{H}^T \boldsymbol{\Omega}\boldsymbol{H}\right)^{-1} \boldsymbol{H}^T \boldsymbol{\Omega}\boldsymbol{Y}\label{eq:wbeta-nlarger}.
\end{align}
Due to the dimension of the matrix to be inverted, if $|\mathcal{V}|> L$, it is more convenient to use \eq{eq:wbeta-nlarger}, while if $|\mathcal{V}| \leq L$ \eq{eq:wbeta-nsmaller} has to be preferred. The output classification for a vector $\boldsymbol{v}$ is then computed as $\argmax_i \left(\boldsymbol{h}(\boldsymbol{v})^T\boldsymbol{B}\right)_i$, where $\boldsymbol{h}(\boldsymbol{v})^T\boldsymbol{B}$ is a vector of WELM scores for each class.

\subsection{WELM based Person Re-Identification}
\label{sec:elm-reid}

To enable person re-identification based on the feature vectors $\boldsymbol{v}$ extracted by the NN, milliTRACE-IR uses the WELM multiclass classifier of \secref{sec:elm-prel}, which is {\it trained at runtime} only when the system has to re-identify a previously seen subject. This is done by sequentially collecting feature vectors from all the subjects seen by the system at operation time, and storing them into the training set $\mathcal{V}$. 

Note that, although an online sequential version of the ELM training process has been proposed in~\cite{huynh2011regularized}, the WELM is trained every time a person has to be re-identified using a batch implementation and including in the training set $\mathcal{V}$ all the subjects seen up to the current time-step $k$. This is because in the online training procedure of~\cite{huynh2011regularized} the number of classes has to be {\it fixed} in advance, while in the considered setup the number of subjects seen by the system may change in time and the Re-Id procedure must be flexible to the addition of new individuals to the training set $\mathcal{V}$. The WELM training and re-identification phases are detailed next and in \alg{alg:reid}.

\subsubsection{Training} the training process is performed at runtime as explained in \secref{sec:elm-prel}, using $L=1,024$ and $\lambda=0.1$. During the normal system operation, the feature vectors obtained from each track are continuously added to set $\mathcal{V}$, storing the corresponding one-hot encoded vectors containing the subjects' identities into matrix $\boldsymbol{Y}$. To reduce the computational burden, the feature extraction step is executed every $5$ time-steps. This is reasonable, as the input sequences to the NN contain $45$ time-steps overall and extracting the features at every time-step would lead to highly correlated, and therefore less informative feature vectors, in addition to entailing a higher computation cost. At time-step $k$, if a subject has to be re-identified, the training procedure of \secref{sec:elm-prel} is executed (lines~$1-4$): the WELM feature vectors $\boldsymbol{H}$ are computed by applying the activation function $\boldsymbol{h}(\cdot)$ to each training vector and the weight matrix $\boldsymbol{\Omega}$ is obtained from \eq{eq:weight-matrix} (lines~$1-3$). The WELM output matrix $\boldsymbol{B}$, is computed using \eq{eq:wbeta-nsmaller} or \eq{eq:wbeta-nlarger} depending on $|\mathcal{V}|$ (line~$4$).

\begin{algorithm}[t!]
\caption{Re-identification mechanism at time $k$.}
\label{alg:reid}
\begin{algorithmic}[1]
\REQUIRE Training set $\mathcal{V}$, track to be re-identified $t^{\rm id}$.
\ENSURE Re-id label of $t^{\rm id}$.
\STATE $\boldsymbol{H}  \leftarrow \left[\boldsymbol{h}^T(\boldsymbol{v}), \forall \, \boldsymbol{v} \in \mathcal{V}\right]$
\STATE $\boldsymbol{Y} \leftarrow $ labels of $\mathcal{V}$
\STATE $\boldsymbol{\Omega} \leftarrow $ \eq{eq:weight-matrix}
\STATE $\boldsymbol{B}  \leftarrow $ \eq{eq:wbeta-nlarger} or \eq{eq:wbeta-nsmaller} depending on $|\mathcal{V}| \lessgtr L $
\STATE $\boldsymbol{\xi}_0 \leftarrow \boldsymbol{0}$
\FOR{$j=1, \dots, W$} 
\STATE $\boldsymbol{v}^{\rm id}_j \leftarrow \mathcal{F}\left(\boldsymbol{Z}_j\right)$
\STATE $\boldsymbol{\xi}_j \leftarrow \left[\boldsymbol{h}^T (\boldsymbol{v}^{\rm id}_j)\boldsymbol{B} + j \boldsymbol{\xi}_{j-1} \right] / (j+1)$
\ENDFOR
\STATE label $\leftarrow \arg \max_i (\boldsymbol{\xi}_W)_i$ 
\end{algorithmic}
\end{algorithm}

\subsubsection{Re-identification} the Re-Id procedure is used to recognize subjects that have been seen by the system and associate them with their temperature measurement and their past movement history in the monitored area.
Denoting by $t^{\rm id}$ the track to be re-identified, the trained WELM processes the NN features of this user, $\boldsymbol{v}^{\rm id}$, as follows:  $\boldsymbol{h}(\boldsymbol{v}^{\rm id})^T\boldsymbol{B}$. Due to the high variability of human movement, rather than considering a single feature vector, milliTRACE-IR computes the cumulative average WELM scores over a time window of length $W$, where the average score at time $j=1,\dots,W$ is referred to as $\boldsymbol{\xi}_j$ (lines~$6-9$). The identity label corresponds to the index of the largest element of $\boldsymbol{\xi}_W$ (line~$10$).

\section{Experimental Results}\label{sec:results}

In this section, the experimental results obtained by testing the system in different indoor environments are presented.

\subsection{Implementation}\label{sec:implementation}
\begin{figure}
	\centering
	\includegraphics[width=5cm]{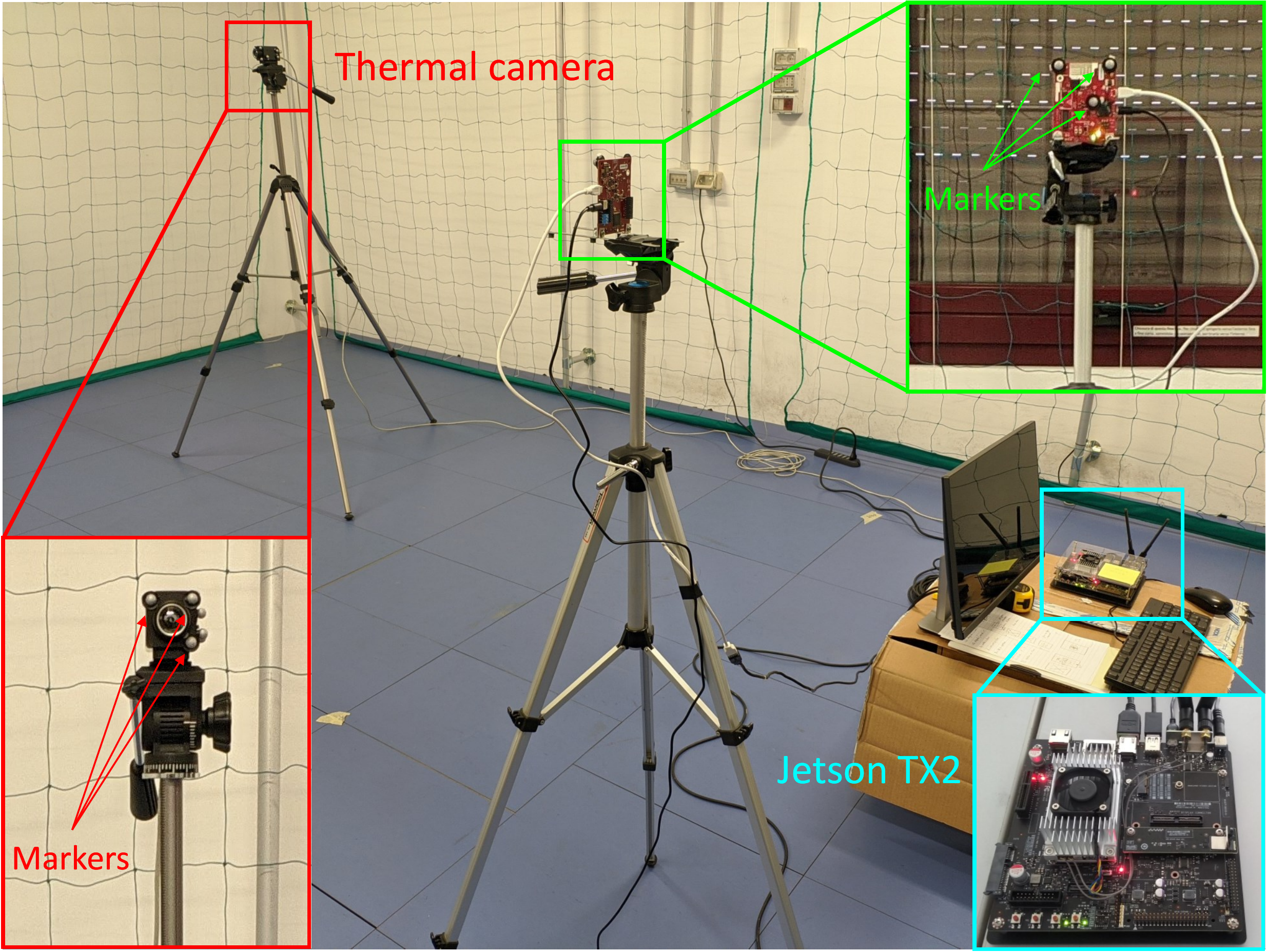}
	\includegraphics[width=2.95cm]{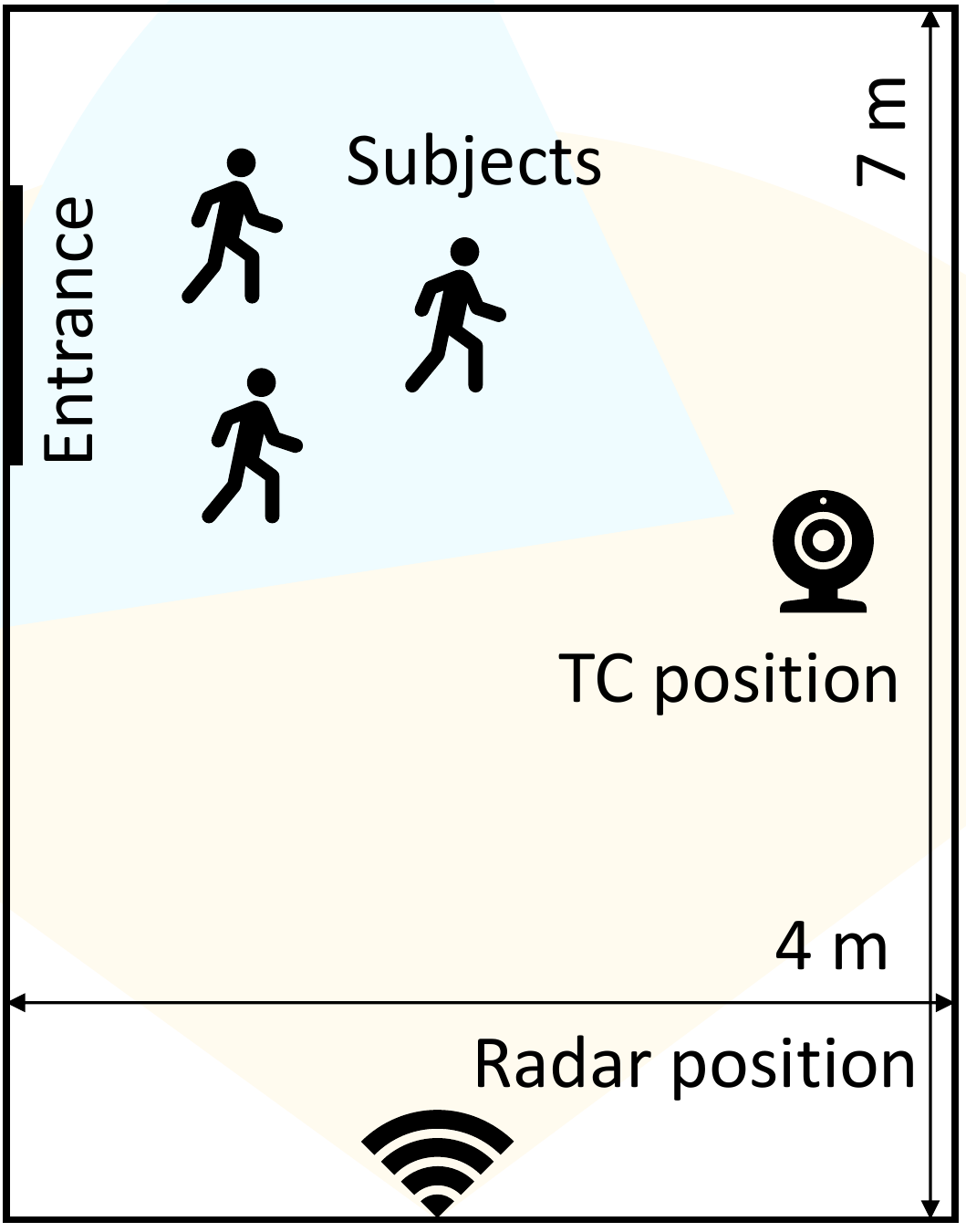}
	\caption{Experimental setup for the data association.}
	\label{fig:exp-setup}
\end{figure}

\noindent \textbf{Hardware.} milliTRACE-IR has been implemented on an NVIDIA Jetson TX2 edge computing device\footnote{https://developer.nvidia.com/embedded/jetson-tx2}, with $8$~GB of RAM and a NVIDIA Pascal GPU. The Jetson TX2 has been connected via USB to a Texas Instruments IWR1843BOOST mmWave radar\footnote{https://www.ti.com/tool/IWR1843BOOST}, operating in the \mbox{$77-81$~GHz} band, and via Ethernet to a FLIR A65 thermal camera\footnote{https://www.flir.it/products/a65/}, as shown in \fig{fig:exp-setup}. The experiments have been performed in real-time at a frame rate of \mbox{$1/\Delta = 15$~Hz}.\\ 
The radar device operates in FMCW mode, using a chirp bandwidth $B=3.07$~GHz, which leads to a range resolution of $c/2B = 4.88$~cm, and $64$ chirps per sequence, obtaining a maximum measurable velocity of $4.77$~m/s and velocity resolution of $14.92$~cm/s.\\
The thermal camera has a $640 \times 512$ focal plane array (FPA), a spectral range of $[7.5, 13]$~$\mu$m, a temperature range of $[-25, 135]^{\circ}$C, a measurement uncertainty of $\pm 5^{\circ}$C, and a noise equivalent temperature difference (NETD) of  $50$~mK. \\
\noindent \textbf{Software.} The system has been developed in Python, using the NumPy, SciPy and OpenCV libraries for the implementation of the tracking phases (for radar and thermal camera) and the proposed data association (\secref{sec:data-assoc}), clustering (\secref{sec:cluster-res}) and re-identification (\secref{sec:elm-reid}) algorithms. Tensorflow and Keras libraries have been used to implement the feature extraction NN (\secref{sec:dl-fextr}). The pre-trained face detector for the thermal images (\secref{sec:tc-ekf}) has been taken from the open-source YOLOFace\footnote{https://github.com/sthanhng/yoloface} implementation.

\subsection{TC and Radar Tracks Association}
\label{sec:assoc-res}
To assess the performance of the radar-TC track association method, experimental tests were conducted in a $7\times 4$~m research laboratory. A motion tracking system including $10$ cameras was used to gather ground-truth (GT) data about the locations of the subjects, by placing markers atop their heads. This camera based tracking system provides 3D localization with millimiter-level precision, for all markers, at a rate of $100$~Hz. The radar and the TC were placed as shown in \fig{fig:exp-setup}. $5$ measurement sequences with $2$ subjects and $9$ sequences with $3$ subjects, all freely entering the  room, were collected. 
The roto-translation matrix $\boldsymbol{\Phi}$ was estimated using a set of markers applied to the devices, while the TC intrinsic matrix $\boldsymbol{\Psi}$ (see \secref{sec:data-assoc}) and the radial distortion coefficients were obtained through the Zhang's method~\cite{zhang2000flexible}, using a sun-heated checkerboard pattern.

An \emph{association} is defined as a specific pairing $i \leftrightarrow j$ of a track $i$ from the radar with a track $j$ from the TC, and a \emph{correct association} as an association for which the two tracks correspond to the same subject. Given a set of tracks, the set of all the correct associations performed by the algorithm is denoted by $\mathcal{A}_{\rm TP}$ (\emph{true positives}), the set of all the associations performed by the algorithm as $\mathcal{A}_{\rm P}$ (\emph{positives}), and the set of all the associations that the algorithm should have performed, based on the GT, as $\mathcal{A}_{\rm R}$ (\emph{relevant}).

To quantify the association performance of the system, define the \emph{precision}, $\mathrm{Pr}=|\mathcal{A}_{\rm TP}|/|\mathcal{A}_{\rm P}|$, and the \emph{recall}, $\mathrm{Rec} = |\mathcal{A}_{\rm TP}|/|\mathcal{A}_{\rm R}|$. 
Using these metrics, the proposed track association method is evaluated by assessing the contribution of each cost component in $A(i,j)$ (see \eq{eqn:cost-function}). The results are reported in \tab{tab:data-assoc-results}, where the row labels $A_x$, $A_d$, and \mbox{$A_x+A_d$} indicate the cost function used. The table also shows the impact of adding the correction coefficient $\rho(K)$ (see \eq{eq:rho}): for the case ``Without $\rho(K)$'', $\rho(K)$ is set to $1$.

\begin{table}[t!] 
	\begin{center}
		\begin{tabular}{lcccc}
			\toprule	
			&\multicolumn{2}{c}{{With $\rho(K)$}}& \multicolumn{2}{c}{{Without $\rho(K)$ }}\\
			\cmidrule(lr){2-3}\cmidrule(lr){4-5}
			&\multicolumn{1}{c}{Pr [\%]}&\multicolumn{1}{c}{Rec [\%]}&\multicolumn{1}{c}{Pr [\%]}&\multicolumn{1}{c}{Rec [\%]}\\
			\cmidrule(lr){2-2}\cmidrule(lr){3-3}\cmidrule(lr){4-4}\cmidrule(lr){5-5}
			$A_x + A_d$ &$\boldsymbol{97.3}$ &$\boldsymbol{97.3}$ &$91.9$ &$91.9$ \\
			$A_x$ only &$91.9$ &$89.2$ &$89.7$ &$94.6$ \\
			$A_d$ only &$92.1$ &$94.6$ &$86.8$ &$89.2$ \\
			\bottomrule
		\end{tabular}
	\end{center}
	\caption{Impact of the components of the cost function. Row labels $A_x$, $A_d$, and \mbox{$A_x+A_d$} indicate, respectively, that only costs $A_x$, $A_d$ or the sum of the two were used in the evaluation. Label ``With $\rho(K)$'' indicate that the corrective term, $\rho(K)$, was used, while label ``Without $\rho(K)$'' means $\rho(K)=1$.}
	\label{tab:data-assoc-results}
\end{table}

As shown, the proposed track association method reliably associates the radar and TC tracks, reaching precision and recall both higher than $97\%$. The joint use of $A_x$, $A_d$ and $\rho(K)$ leads to improvements of up to $11\%$ and $8\%$ for the precision and recall metrics, respectively.

\subsection{Temperature Screening}\label{sec:temp-screen}

\begin{figure}[t!]
		\centering
		\subcaptionbox{Comparison between with (\emph{Corr. temp.}) and without (\emph{Raw temp.}) distance-based correction. The triangles show the mean values. \label{fig:temp-methods}}[0.48\columnwidth]{\includegraphics[height=4.5cm]{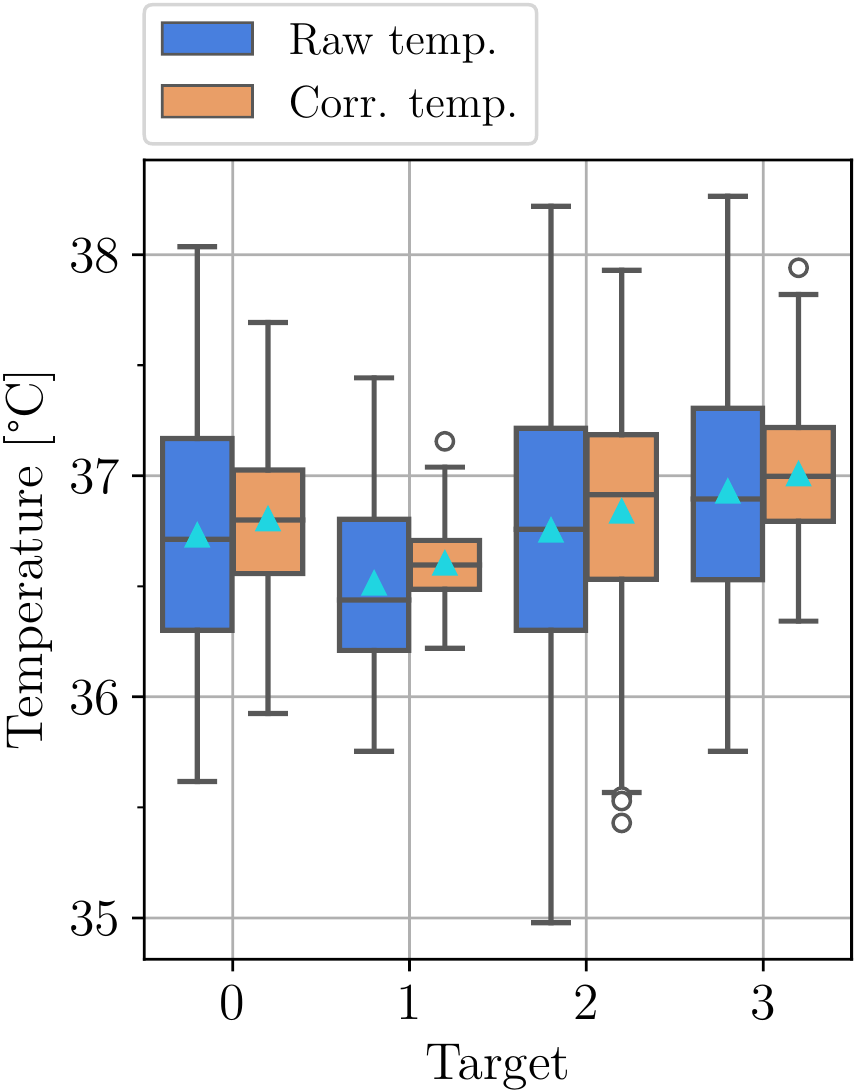}}
		\subcaptionbox{Comparison between the estimated temperatures and the true temperatures. The error bars represent the standard deviations. \label{fig:temp-estimate}}[0.48\columnwidth]{\includegraphics[height=4.5cm]{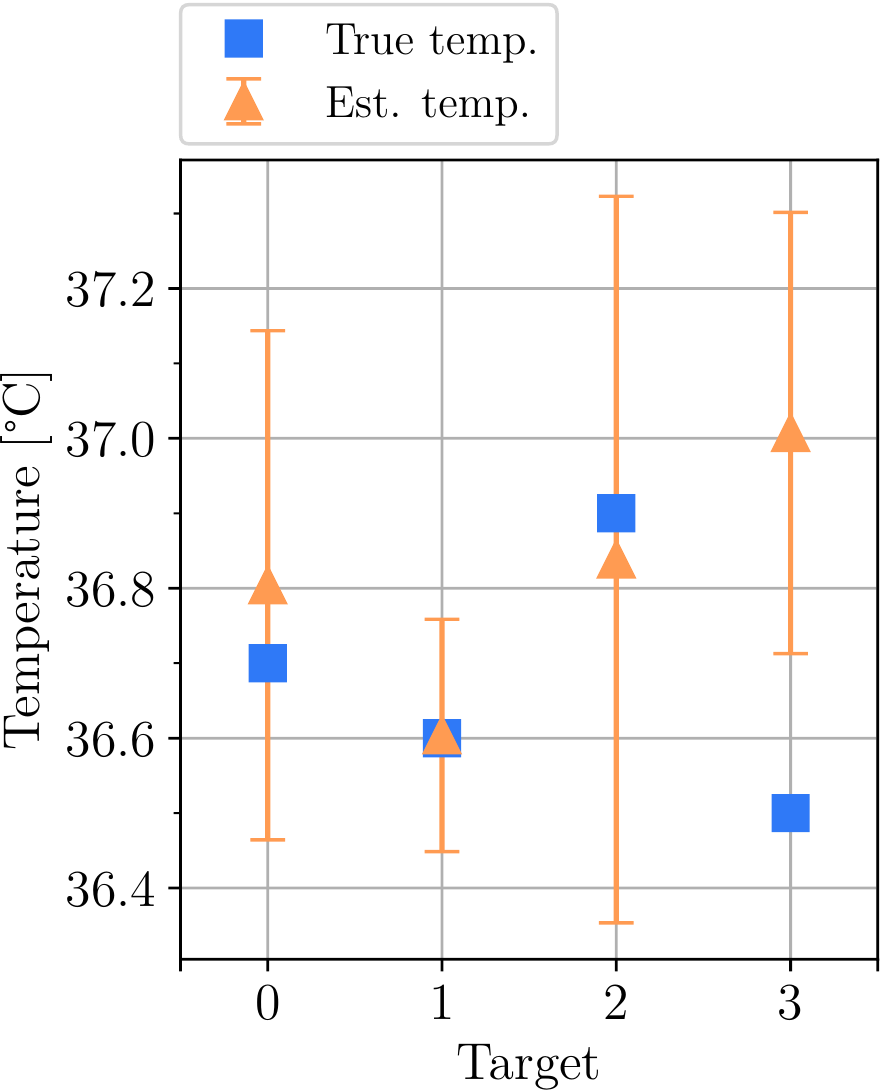}}
		\caption{Results of the temperature screening.}
		\label{fig:temp}
\end{figure}

\begin{figure}[t]
	\begin{center}   
		\includegraphics[width=0.85\columnwidth]{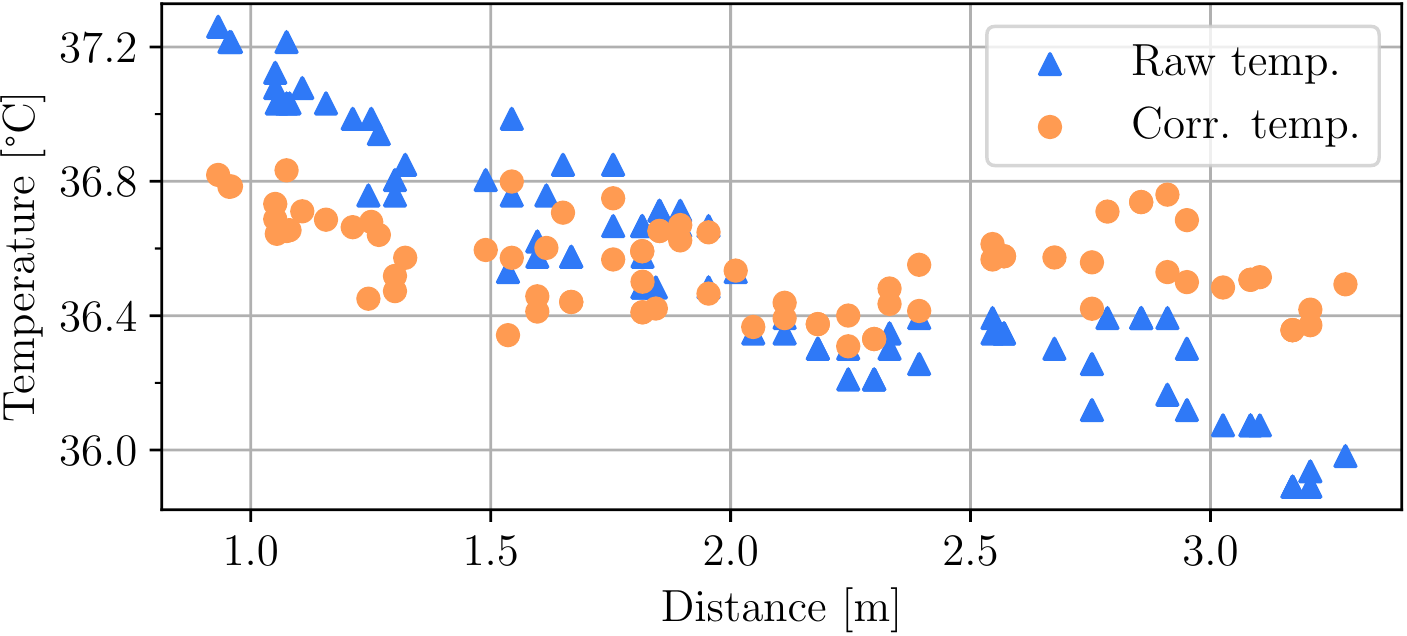}
		\caption{Temperature measurements from a subject moving in front of the TC with (\emph{Corr. temp.}) and without (\emph{Raw temp.}) distance-based correction.}
		\label{fig:temp-correction-example}
	\end{center}
\end{figure}
Remarkably, the proposed temperature screening method does not require people to stand in front of the TC sensor, but estimates their temperature as they move within the FoV of the TC. In order for the method to return accurate temperature measurements, the subject' frontal face should be captured by the TC for a minimum time duration. For this reason, it is advisable to place the TC near a point of passage, e.g., in proximity of an entrance. The temperature screening method was tested on $4-7$ sequences of $\sim10$~s each were collected from $4$ different individuals moving within $3.5$~m from the TC. Each subject was tested at a different time of the day, to gauge the effects of the changing (thermal) environmental conditions, and of a possible concept drift (e.g., heating) of the TC after a long period of operation. Furthermore, as explained in \secref{sec:temp-est}, a linear function $\alpha(\cdot)$ was fit to compensate for the influence of the distance on the measures.

To evaluate the benefit brought by the correction based on the targets' distance, in \fig{fig:temp-methods}, the results obtained with (\emph{Corr. temp.}) and without (\emph{Raw temp.}) the correction are compared. Since the TC is intrinsically subject to a bias, to facilitate the comparison of the measures, in the \emph{Raw temp.} case only this bias is corrected, assuming a constant target distance of $2$~m and multiplying each measured temperature by \mbox{$\alpha(2) = a_0 + 2 a_1$}. The full method (\emph{Corr. temp.}), instead, uses the rescaled average estimate, as per \eq{eq:corrected-temp}. The box-plot shows that the range of the corrected temperatures is significantly reduced (for these experiments, the true temperature is constant), demonstrating the efficacy of the proposed correction plus averaging approach. As an illustrative example, \fig{fig:temp-correction-example} shows the impact of the distance-based correction on data measurements from a subject moving in front of the TC.


\begin{table}[t!] 
	\begin{center}
		\begin{tabular}{lcccc}
			\toprule
			& Mean [°C] & $\pm$ std [°C] & True temp. [°C] & Error [°C] \\
			\cmidrule(lr){2-5}
			Target $0$     & $36.8$& $0.340$& $36.7$& $0.104$ \\
			Target $1$     & $36.6$& $0.155$& $36.6$& $0.004$ \\
			Target $2$     & $36.8$& $\boldsymbol{0.485}$& $36.9$& $-0.062$ \\
			Target $3$     & $37.0$& $0.294$& $36.5$& $\boldsymbol{0.507}$ \\
			\bottomrule
		\end{tabular} 		
	\end{center}
	\caption{Results of the temperature estimation and comparison with respect to the true values for the $4$ targets. The worst cases are highlighted.} 
	\label{tab:temp-estimate}
\end{table}

\fig{fig:temp-estimate} compares the temperature estimates from milliTRACE-IR and the true temperatures measured with a contact thermometer. The numerical results are reported in \tab{tab:temp-estimate}, where the worst cases are reported in bold fonts.
Mean temperatures are estimated with a maximum standard deviation from the mean smaller than $0.5$~°C and a maximum absolute error with respect to the true temperature of about $0.5$~°C.
Note that only one of the subjects in \fig{fig:temp-estimate} exhibits this maximum error (subject $3$), while the absolute error for the others remains within $0.1$~°C. These errors descend from the fact that the environmental conditions and the heating of the thermal camera affect the measurements in an unpredictable way, modifying the \emph{bias} of the fitting function. Notwithstanding, the thermal screening capability of \mbox{milliTRACE-IR} is significantly better than that of existing approaches, see \secref{sec:comparison}. Also, some improvements are possible by, e.g., applying a correction based on an external reference, such as a piece of material instrumented with a contact thermometer and located within the field of view of the TC, or monitoring the statistics of the people's temperature (mean $\mu$ and standard deviation $\sigma$) to detect anomalous samples within such empirical distribution. For instance, an alarm could be raised for those subjects whose temperature is greater than $\mu + c \times \sigma$, for a user-defined threshold $c$. This would allow the system to continuously and autonomously adapt to different operating conditions. 

\subsection{Positioning and Social Distance Monitoring}
\label{sec:dist-res}

To evaluate the performance of the radar tracking system in estimating the position of the targets and the inter-subject's distance, tests were conducted in the $7\times 4$~m research laboratory described in \secref{sec:assoc-res}. A total of $7$ sequences of duration $~10-15$~s were collected, each with $3$ subjects moving freely in the room, along with their GT locations obtained from the motion tracking system. The root mean squared error (RMSE) between the mmWave radar estimated locations and the GT is used as a performance metric. Moreover, the inter-subject distances were measured, considering all the possible combinations of the three subjects and leading to a total of $21$ inter-subject distances across all the recorded sequences.

\begin{figure}
\centering
\includegraphics[width=0.8\columnwidth]{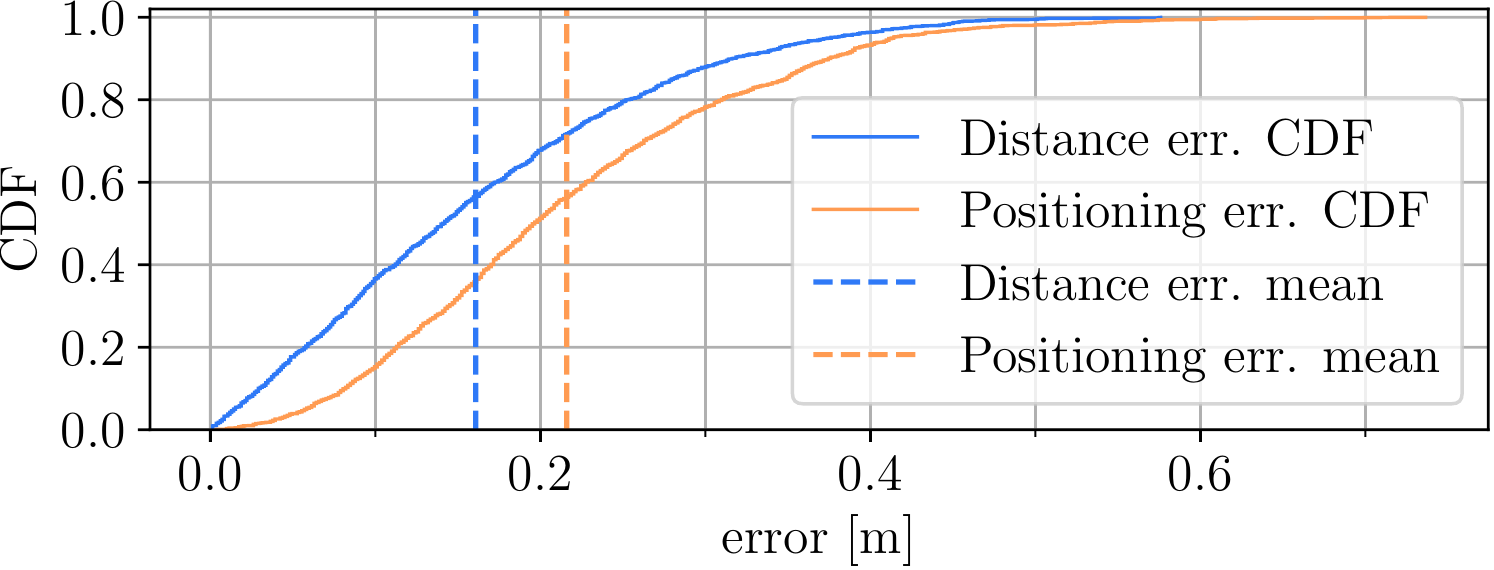}
\caption{CDF of the absolute error between the true (ground truth) and the estimated subject's position / inter-subject's distance, as measured by the radar tracking system. The dashed lines denote the mean error.}
\label{fig:cdf-error}
\end{figure}

The cumulative distribution functions (CDF) of the absolute error between the ground truth and the estimated subject's position/inter-subject distance, as measured by the radar tracking system, is shown in \fig{fig:cdf-error}, along with the corresponding mean values. The numerical results are provided in \tab{tab:pos-results}. The radar system achieves an absolute \emph{positioning} error within $0.3$~m in $80\%$ of the cases. For the inter-subject \emph{distance}, the error remains within $0.25$~m in $80\%$ of the cases.

\begin{table}[t!] 
\begin{center}
\begin{tabular}{lcccc}
	\toprule
	& Mean [m] & $\pm$ std [m] & Frames & Time [s] \\
	\cmidrule(lr){2-5}
	Position RMSE      &$0.216$   & $0.115$       & 1448   & ~97 \\
	Subj. distance RMSE&$0.161$   & $0.112$       & 1153   & ~77 \\
	\bottomrule
\end{tabular} 		
\end{center}
\caption{RMSE of the subject's position and of the inter-subject's distance estimated by the radar sensor, computed against the GT.} 
\label{tab:pos-results}
\end{table}

\begin{table}[t!] 
\begin{center}
\begin{tabular}{lcccc}
	\toprule	
	&\multicolumn{2}{c}{{\bf milliTRACE-IR}}& \multicolumn{2}{c}{{\bf DBSCAN }}\\
	\cmidrule(lr){2-3}\cmidrule(lr){4-5}
	&\multicolumn{1}{c}{$r_{\rm cl}$ $[\%]$}&\multicolumn{1}{c}{corr. tracked}&\multicolumn{1}{c}{$r_{\rm cl}$ $[\%]$}&\multicolumn{1}{c}{corr. tracked}\\
	\cmidrule(lr){2-2}\cmidrule(lr){3-3}\cmidrule(lr){4-4}\cmidrule(lr){5-5}
	$2$ sub. parallel&$90.7$&$\checkmark$&$46.5$&$\times$\\
	$2$ sub. crossing&$87.9$&$\checkmark$&$59.6$&$\times$\\
	$2$ sub. close&$89.9$&$\checkmark$&$69.7$&$\checkmark$\\
	$3$ sub. parallel&$92.3$&$\checkmark$&$65.3$&$\checkmark$\\
	$3$ sub. crossing&$83.7$&$\checkmark$&$73.5$&$\times$\\
	\bottomrule
\end{tabular} 		
\end{center}
\caption{Ratio $r_{\rm cl}$ between the number of frames in which the different subjects are correctly separated and the total number of frames, using the proposed method and DBSCAN. Symbols ``$\checkmark$'' and ``$\times$'' denote success and failure of the tracking step, respectively.} \label{tab:clust-res}
\end{table}

\begin{figure*}[t!]
\begin{center}   
\centering
\subcaptionbox{$1$~min. training data. \label{fig:reid-1min}}[5.5cm]{\includegraphics[width=5.5cm]{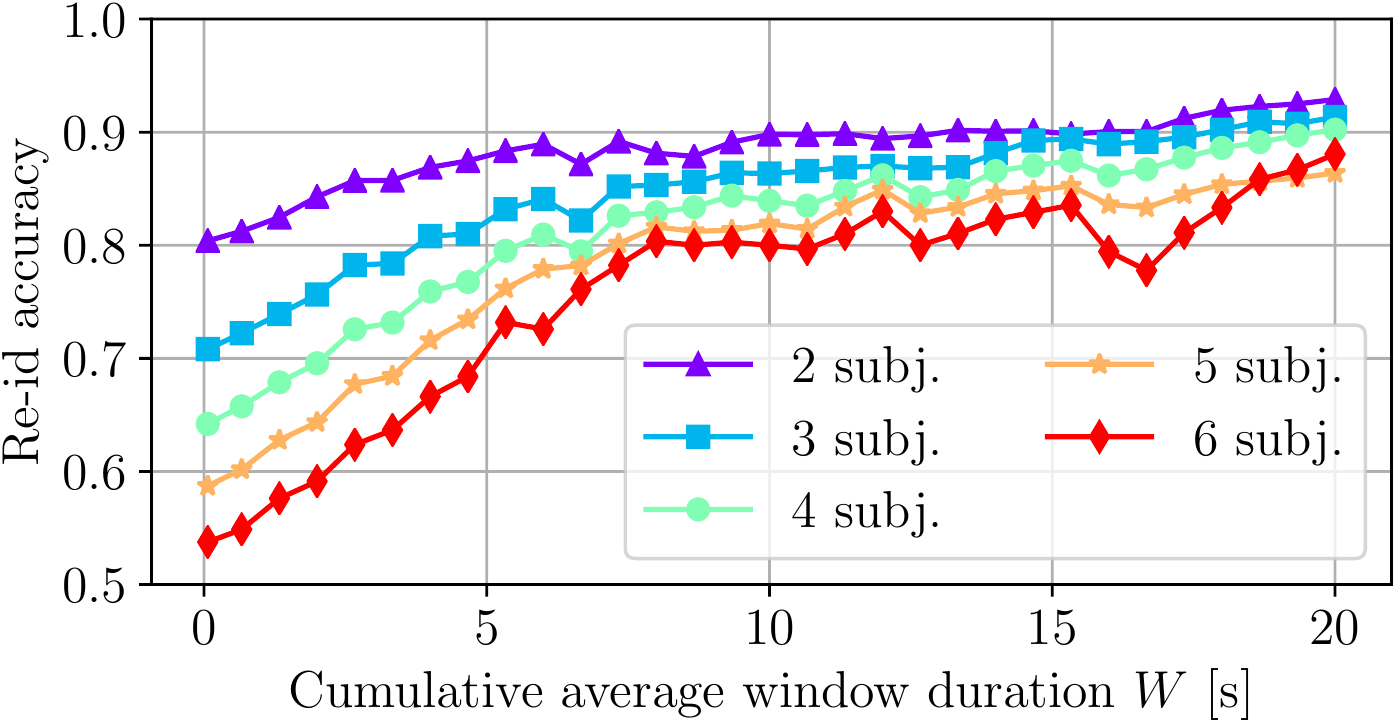}}
\subcaptionbox{$3$~min. training data.\label{fig:reid-3min}}[5.5cm]{\includegraphics[width=5.5cm]{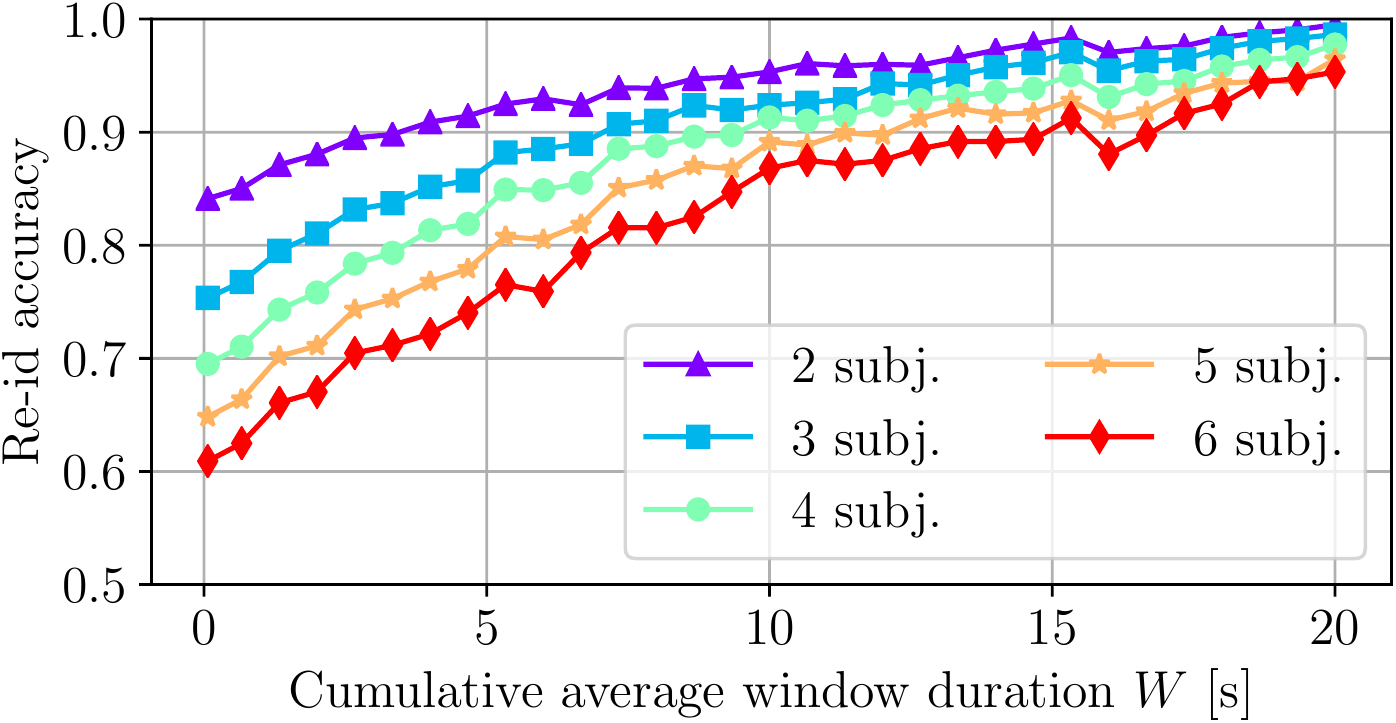}}
\subcaptionbox{Imbalanced training data.\label{fig:reid-unb}}[5.5cm]{\includegraphics[width=5.5cm]{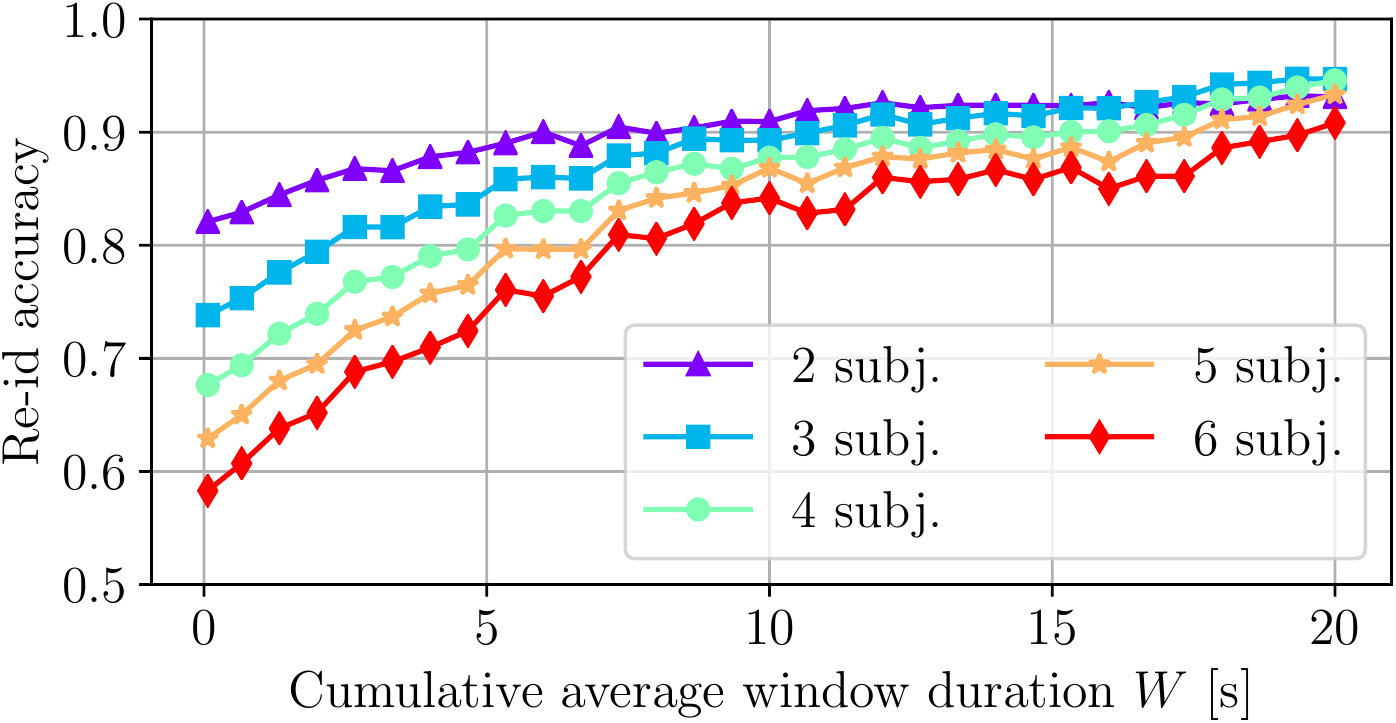}}
\caption{Re-identification accuracy results. In (a) and (b) the re-identification algorithm is used with $1$ and $3$ minutes of training data per subject, respectively. In (c), $1$ minute of training data was used for a randomly selected subset containing half of the subjects, while $4$ minutes were used for the remaining half.}
\label{fig:reid}
\end{center}
\vspace{-0.3cm}
\end{figure*}
\subsection{Effectiveness of the Improved Clustering Technique} 
\label{sec:clust-res}

To evaluate the improvement brought by the proposed clustering method over the standard DBSCAN, both algorithms were tested on specific measurement sequences with subjects moving within $1$~m from one another. To quantify the clustering performance, the correct clustering ratio, $r_{\rm cl}$, is used. This metric represents the fraction of frames in which the clusters belonging to the different subjects are correctly separated. The results of this evaluation are summarized in \tab{tab:clust-res}. The evaluation is conducted on sequences with $2$ and $3$ individuals \textit{(i)} walking along parallel paths with the same velocity and at a distance between $0.5$~m and $0.8$~m (\textit{parallel}), \textit{(ii)} walking along crossing paths, with subjects coming as close as $0.2$~m from one another (\textit{crossing}) and \textit{(iii)} staying still and moving arms at an inter-subject distance of approximately $0.8$~m (\textit{close}). The proposed clustering algorithm led to a large improvement (up to $44$~$\%$) in terms of $r_{\rm cl}$ metric with respect to DBSCAN. In addition, for $3$ of the $5$ test sequences, DBSCAN led to failures in the tracking process, either merging the tracks of different subjects, or failing to detect some of them, while milliTRACE-IR correctly tracked all the subjects in all cases.

\subsection{Person Re-Identification} \label{sec:reid-res}


The proposed WELM based Re-Id algorithm was evaluated on a set of mmWave radar measurements from $6$ individuals who were \textit{not} included among the $16$ subjects used to train the feature extraction NN. The tests were conducted in a $12\times 3$~m research lab, with furniture that made the evaluation challenging. The training data contains $4$~minutes of measurements ($3,600$ radar frames) while over $1$~minute of measurements per subject ($1,000$ frames) was used as test data. In both the training and the test data, the individuals walked freely in the room. The radar position was changed for each test to gauge the impact of varying the radar point-of-view. 

\noindent \textbf{Re-Id accuracy.}
The Re-Id accuracy as a function of $W$ (see \alg{alg:reid}) is shown in \fig{fig:reid-1min} and \fig{fig:reid-3min}. The curves of these plots are obtained averaging the results of $20$ different WELM initializations, and all the possible combinations of the considered number of subjects (from $2$ to $6$) over the $6$ total individuals. As expected, the Re-Id performance increases with an increasing inference time (larger $W$) and with the length of the training sequences: the accuracy gain is about $10\%$ by going from $1$-minute (\fig{fig:reid-1min}) to $3$~minutes (\fig{fig:reid-3min}) long training sequences. Also, milliTRACE-IR reaches high Re-Id accuracy using \mbox{$W \geq 15$~s} and the detrimental effect of increasing number of subjects to be classified is greatly reduced using larger values of $W$, as accumulating the WELM scores over longer time windows increases the robustness of the WELM decision. Overall, the accuracy of the proposed method is higher than $95\%$ in all cases, only using $3$~minutes of training data per subject and $W=20$~s, which are reasonable in practice. The worst-case ($3$~minutes of training data for $6$ subjects) WELM training time, on the ARM Cortex-A57 processor of the Jetson TX2 device, took $2.98\pm 0.015$~s. 

\noindent \textbf{Impact of imbalanced training data.}
As shown in \fig{fig:reid-unb}, the effect of imbalanced training data is successfully mitigated by the sample weighting strategy of \eq{eq:weight-matrix}. In this evaluation, the WELM was trained with $1$~minute of data for a randomly selected subset containing half of the subjects and $4$~minutes for the remaining half. 


\noindent \textbf{Improvement over a baseline.} \tab{tab:elm-base-comp} compares the WELM to a baseline classification method widely used in camera-based person Re-Id~\cite{ye2021deep} that, unlike milliTRACE-IR, does not learn a similarity score based on the actual distribution of the feature vectors at operation time. The baseline algorithm collects the training feature vectors along with the corresponding labels and computes the \textit{centroid} of each class $m$ in the NN feature space, denoted by $\boldsymbol{c}_m$. To re-identify a subject, the \textit{cosine similarity} (CS) between his/her feature vectors, $\boldsymbol{v}$, and the centroid of each class $m$ is computed, obtaining a similarity score $s_m = \boldsymbol{c}_m^T\boldsymbol{v}/ (||\boldsymbol{c}_m||_2\times||\boldsymbol{v}||_2)$, and the classification is performed taking $\argmax_{m}s_m$. 
The WELM outperforms the baseline scheme in all the tests, see \tab{tab:elm-base-comp}. The performance gap is significant for little training data (up to $16\%$ improvement), small windows and imbalanced training sets. 

\begin{table}[t!] 
\begin{center}
\begin{tabular}{lcccccc}
\toprule	
&\multicolumn{3}{c}{{\bf WELM}}& \multicolumn{3}{c}{{\bf CS baseline }}\\
\cmidrule(lr){2-4}\cmidrule(lr){5-7}
&\multicolumn{1}{c}{$1$~min. }&\multicolumn{1}{c}{$4$~min.}
&\multicolumn{1}{c}{imb.}&\multicolumn{1}{c}{$1$~min. }&\multicolumn{1}{c}{$4$~min. }&\multicolumn{1}{c}{imb.}\\
\cmidrule(lr){2-2}\cmidrule(lr){3-3}\cmidrule(lr){4-4}\cmidrule(lr){5-5}\cmidrule(lr){6-6}\cmidrule(lr){7-7}
$W=0$~s&$53.8$&$60.9$&$58.3$&$44.6$&$49.5$&$51.6$\\
$W=10$~s&$80.0$&$86.8$&$84.2$&$63.9$&$77.7$&$80.6$\\
$W=20$~s&$88.6$&$95.3$&$90.8$&$72.2$&$88.8$&$86.9$\\
\bottomrule
\end{tabular} 		
\end{center}
\caption{Re-Id accuracies obtained by the WELM and the CS baseline on $6$ subjects using $1$ and $4$~minutes balanced training sets, and an imbalanced training set. The cumulative average window $W$ is set to $0$~s (a single test feature vector is used), $10$~s or $20$~s.} \label{tab:elm-base-comp}
\end{table}


\subsection{Comparison with existing approaches}
\label{sec:comparison}
\begin{table}[t!]
\setlength{\tabcolsep}{2.5pt}
\begin{center}
\begin{tabular}{lccc}
\toprule
& \textbf{milliTRACE-IR} & \thead{\textbf{\"{U}lrich} \cite{ulrich2018person}} & \thead{\textbf{Savazzi} \cite{savazzi2020processing}} \\
\cmidrule(lr){2-2} \cmidrule(lr){3-3} \cmidrule(lr){4-4}
\makecell[l]{\tabentry{Positioning range} \tabentry{RMSE [m]}}          & $0.19$     & $\times$ & $0.45$ \\
\makecell[l]{\tabentry{Interpersonal dist.} \tabentry{RMSE [m]}}         & $0.17$    & $\times$ & $0.5$* \\
\makecell[l]{\tabentry{Positioning angle} \tabentry{RMSE [°]}}          & $3.1$      & $\times$ & $7.0$ \\
\makecell[l]{\tabentry{Thermal screening} \tabentry{RMSE [$^{\circ}$C]}} & $0.13$     & $\times$ & $0.45$ \\
\makecell[l]{\tabentry{Thermal screening} \tabentry{range [m]}}          & $3.5$       & $\times$ & $1.1$ \\
\makecell[l]{\tabentry{Per-frame assoc.} \tabentry{Pr [\%]}}             & $98.6$      & $25.2$   & n.a. \\
\makecell[l]{\tabentry{Per-frame assoc.} \tabentry{Rec [\%]}}            & $98.5$       & $78.4$   & n.a. \\
\makecell[l]{\tabentry{Re-identification} \tabentry{acc. [\%]}}          & $\approx90$  & $\times$ & $\times$ \\
\bottomrule
\end{tabular} 		
\end{center}
\caption{Comparison of milliTRACE-IR with the works from \"{U}lrich et al. \cite{ulrich2018person} and Savazzi et al. \cite{savazzi2020processing}. Symbols ``$\times$" and ``n.a." denote, respectively, that the task is not tackled or that there is no available result for the considered quantity in the original papers. The symbol ``*" is used to highlight that the value is not an RMSE value but the minimum interpersonal distance threshold considered in~\cite{savazzi2020processing}. 
}
\label{tab:comparison}
\end{table}

In this section a comparison between milliTRACE-IR and available methods from the literature is provided.
To the best of the authors' knowledge, only two works exploit both mmWave radars and thermal cameras to perform human sensing and/or temperature screening, namely, the works from \"{U}lrich et al.~\cite{ulrich2018person} and Savazzi et al.~\cite{savazzi2020processing}. Since none of the two tackles all the points that milliTRACE-IR addresses, they are here considered, separately, to compare different aspects. The data association strategy is compared with that proposed in~\cite{ulrich2018person}, while~\cite{savazzi2020processing} is used to compare the positioning, distance monitoring, and temperature screening parts.
In \tab{tab:comparison}, symbols ``$\times$" and ``n.a." denote, respectively, that the task is not tackled or that no specific result is provided in the corresponding work.

\noindent
\textbf{Data association.} 
In~\cite{ulrich2018person} (\"{U}lrich et al.), people are detected in thermal images by applying the Viola and Jones algorithm~\cite{viola2001robust} to detect the upper bodies of the subjects in the environment. The distance between the TC and each subject is roughly retrieved from the dimension of the bounding box enclosing the upper body of each person, similarly to what milliTRACE-IR does with faces. The TC detections are then associated, on a frame basis, with range measurements obtained with a mmWave radar by minimizing a Gaussian-shaped association cost. This cost provides an estimate of the probability that the corresponding association is correct, based on the difference between the distance estimates by the TC and by the mmWave radar.
This data association method has been implemented and tested on the dataset of~\secref{sec:assoc-res}, comparing it to the data association strategy of milliTRACE-IR. For a fair comparison, the YOLOv3 detector has been used in place of the Viola and Jones algorithm, as, besides providing superior performance, it is the same detector used by milliTRACE-IR. This guarantees that any difference in the data association results is only due to the data association strategy.
At every time frame, each bounding box has been associated with the radar detection yielding the highest association probability, which corresponds to the smallest difference in the two distance estimates. The main differences between the approach in~\cite{ulrich2018person} and that of milliTRACE-IR are that, in~\cite{ulrich2018person}: \emph{(i)} the association is per-frame and not per-track, \emph{(ii)} the estimated distance from the TC is the only feature considered for the association, and \emph{(iii)} the Hungarian algorithm is not used, so different bounding boxes can be, erroneously, associated with the same radar detection.
Numerical results for the precision (``Pr'') and recall (``Rec'') metrics are presented in \tab{tab:comparison}. Since the association technique of~\cite{ulrich2018person} performs a per-frame association, the table shows the per-frame performance of milliTRACE-IR, computed by counting the number of frames that are correctly classified using milliTRACE-IR's per-track association algorithm. From these results, it can be seen that milliTRACE-IR performs notably better in associating mmWave radar with TC human detections. The largest improvement is brought by the combination of milliTRACE-IR \textit{per-track} association paradigm with the Hungarian algorithm, which effectively filters out ghost tracks and spurious detections which often occur in real world scenarios, significantly boosting the robustness of the scheme.

\noindent
\textbf{Positioning, distancing, and temperature screening.} In~\cite{savazzi2020processing} (Savazzi et al.), people localization, interpersonal distance monitoring, and temperature screening are addressed using thermopiles and mmWave radars.
Since in~\cite{savazzi2020processing} the data association strategy is not disclosed, a comparison is here provided only for the previously mentioned tasks. 
In the paper, positioning performance is evaluated in terms of range (radial distance) and angular RMSEs. Numerical values for these metrics are given in \tab{tab:comparison} considering the dataset of \secref{sec:dist-res} for milliTRACE-IR  and the (average) values from Tab.~II of~\cite{savazzi2020processing} for their algorithm.

In the same work, interpersonal distance monitoring is obtained by dividing the monitored area into a regular grid, whose cells have a side length of $0.5$~m. The system is able to distinguish subjects occupying adjacent cells, which are considered to be violating the minimum interpersonal distance of $1$~m, thus raising an alarm. For this reason, the resolution of the method of~\cite{savazzi2020processing} is $0.5$~m in the best case (a lower bound for the interpersonal distance estimation error). In \tab{tab:comparison}, this value is reported alongside the RMSE of milliTRACE-IR in measuring interpersonal distances, marking the former with a ``*" symbol, to highlight that it is not an RMSE.\\
Thermal screening performance comparisons are also presented in \tab{tab:comparison}, where ``Thermal screening range [m]" refers to the maximum distance at which the tests were carried out. milliTRACE-IR performs better than~\cite{savazzi2020processing} in all the considered tasks, showing a larger monitoring range and more accurate body temperature estimates. In addition, \mbox{milliTRACE-IR} combines these monitoring capabilities with a robust data association strategy and with the capability to re-identify subjects when moving through different areas.

\section{Concluding Remarks}
\label{sec:conclusion}

This work presents the design and implementation of \mbox{milliTRACE-IR}, the first system combining high resolution mmWave radar devices and infrared cameras to perform non-invasive joint temperature screening and contact tracing in indoor spaces. This system uses thermal cameras to infer the temperature of the subjects, achieving measurement errors within $0.5$~°C, and mmWave radars to infer their spatial coordinates, by successfully locating and tracking subjects that are as close as $0.2$~m apart. This is possible thanks to improvements along several lines, such as the association of the thermal camera and radar tracks from the same subject, along with a novel clustering algorithm combining density-based and Gaussian mixture methods to separate the radar reflections coming from different subjects as they move close to one another. Moreover, milliTRACE-IR performs contact tracing: a person with high body temperature is reliably detected by the thermal camera sensor and subsequently traced across a large indoor area in a non-invasive way by the radars. When entering a new room, this subject is re-identified among several other individuals with high accuracy ($95\%$), by computing gait-related features from the radar reflections through a deep neural network and using a weighted extreme learning machine as the final re-identification tool.

Future research includes improvements of the re-identification mechanism for contact tracing. \mbox{milliTRACE-IR} uses an offline pre-training phase for the neural network based feature extraction block, which is performed on a dataset including several subjects. This step could be possibly removed by exploring recent developments in self-supervised learning, thus greatly enhancing the generality and usability of the system. Self-supervised learning could allow the automatic training of the gait feature extractor in a fully online fashion, by exploiting the human movement traces of opportunity that are gathered during system operation. Moreover, a large scale implementation of \mbox{milliTRACE-IR}, featuring tens of mm-Wave radars distributed across a large and crowded indoor environment is another interesting research direction. In such setup, it is key to develop data fusion and collaborative sensing algorithms for radars with overlapping fields of view, towards providing improved resilience to occlusions and better human tracking performance.

\appendices


\bibliography{biblio}
\bibliographystyle{ieeetr}

%


\begin{IEEEbiography}[{\includegraphics[width=1in,height=1.25in,clip,keepaspectratio]{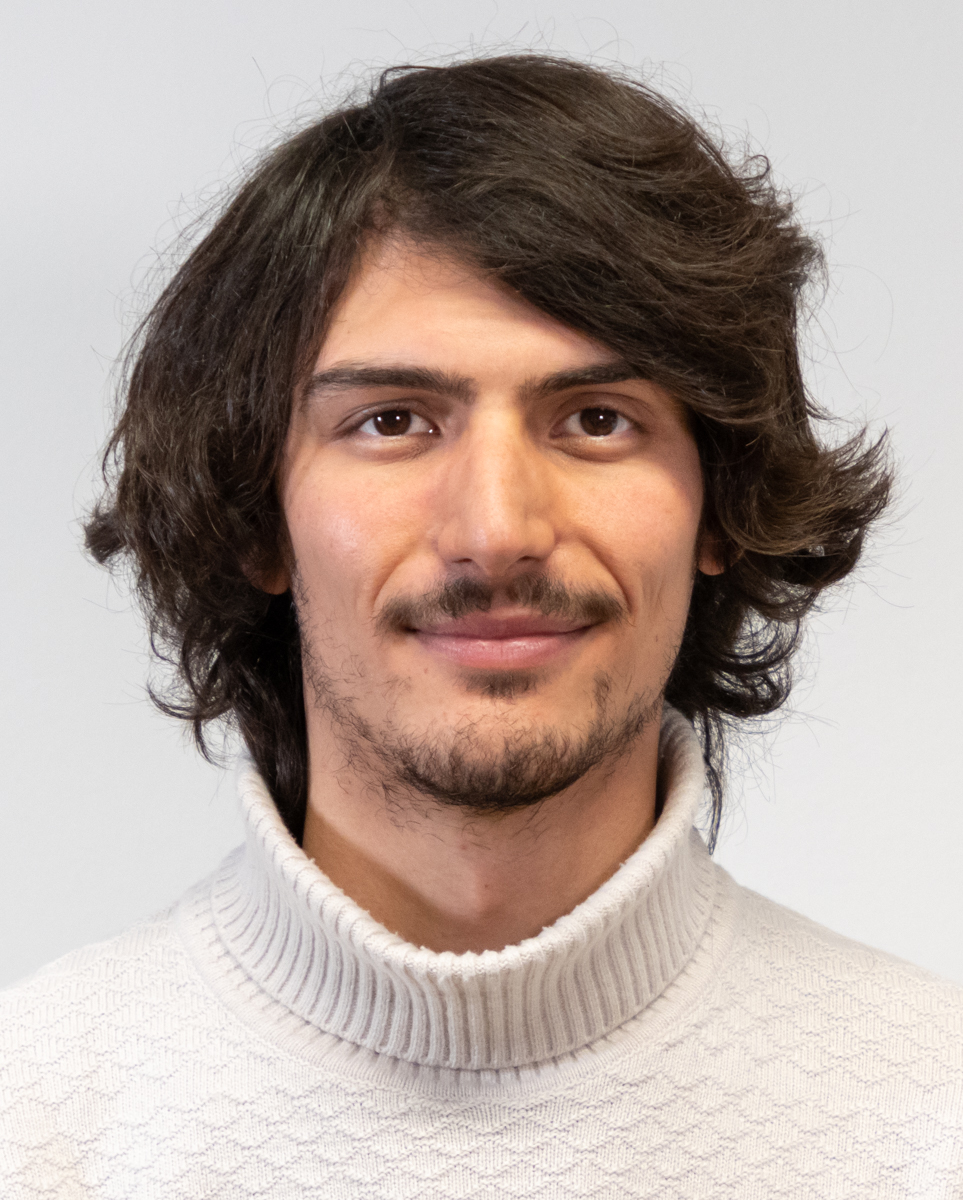}}]%
	{Marco Canil} (S'20) received the B.Sc. degree in Information Engineering in 2019 and the M.Sc. degree in ICT for Internet and Multimedia Engineering in 2020, at the University of Padova (UNIPD), Italy.
	He is currently pursuing a Ph.D. in Information Engineering with the SIGNET research group of the Department of Information Engineering (DEI), in the same university.
	His research interests include machine learning, signal processing, sensor fusion, and remote sensing with mmWaves.
\end{IEEEbiography}

\begin{IEEEbiography}[{\includegraphics[width=1in,height=1.25in,clip,keepaspectratio]{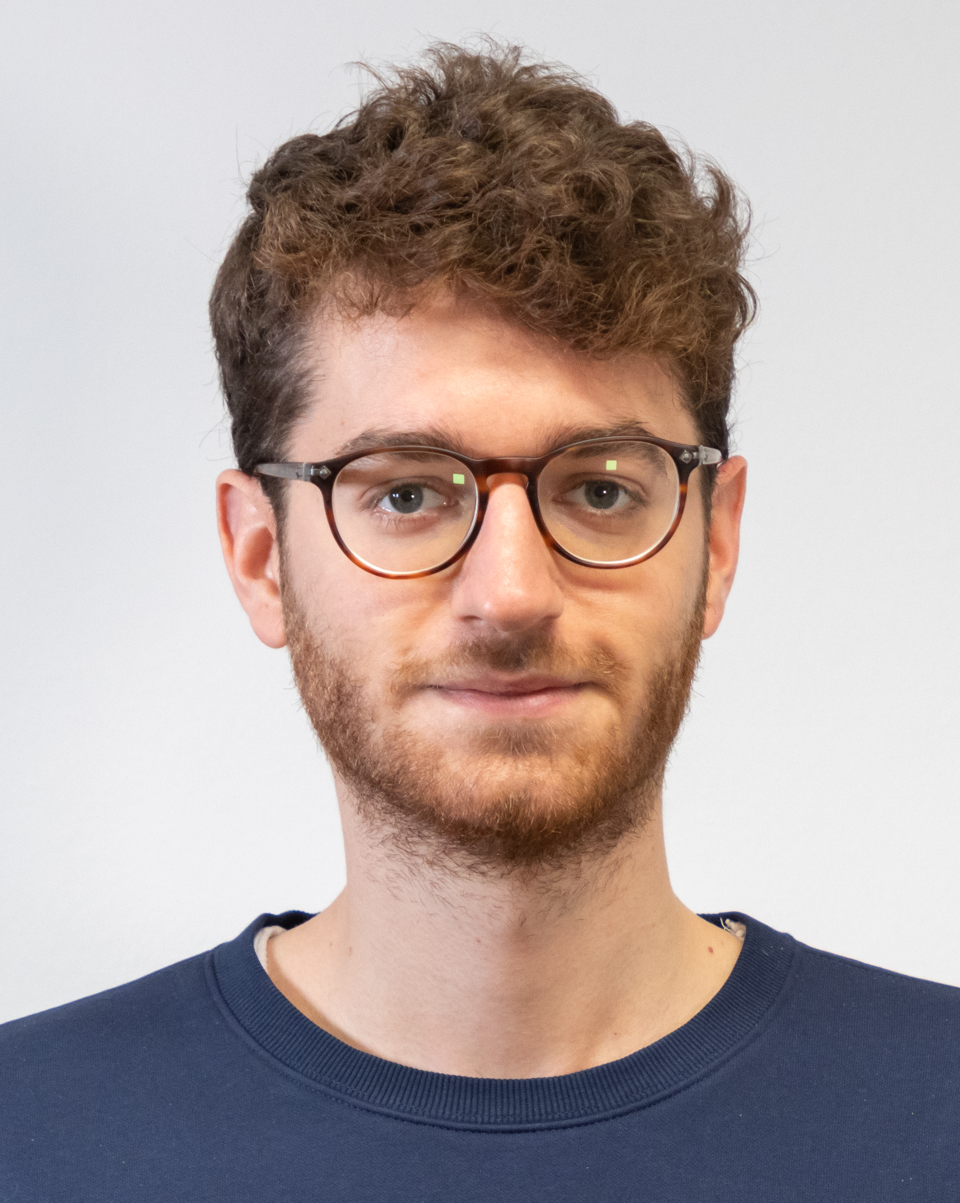}}]%
	{Jacopo Pegoraro} (S'20) received the B.Sc. degree in information engineering and the M.Sc. degree in ICT for Internet and Multimedia engineering from the University of Padova, Padua, Italy, in 2017 and 2019, respectively. He is currently pursuing the
	Ph.D. degree with the SIGNET Research Group, Department of Information Engineering, in the same University. His research interests include signal processing, sensor fusion and machine learning with applications to mmWave sensing and integrated sensing and communication solutions.
\end{IEEEbiography}


\begin{IEEEbiography}[{\includegraphics[width=1in,height=1.25in,clip,keepaspectratio]{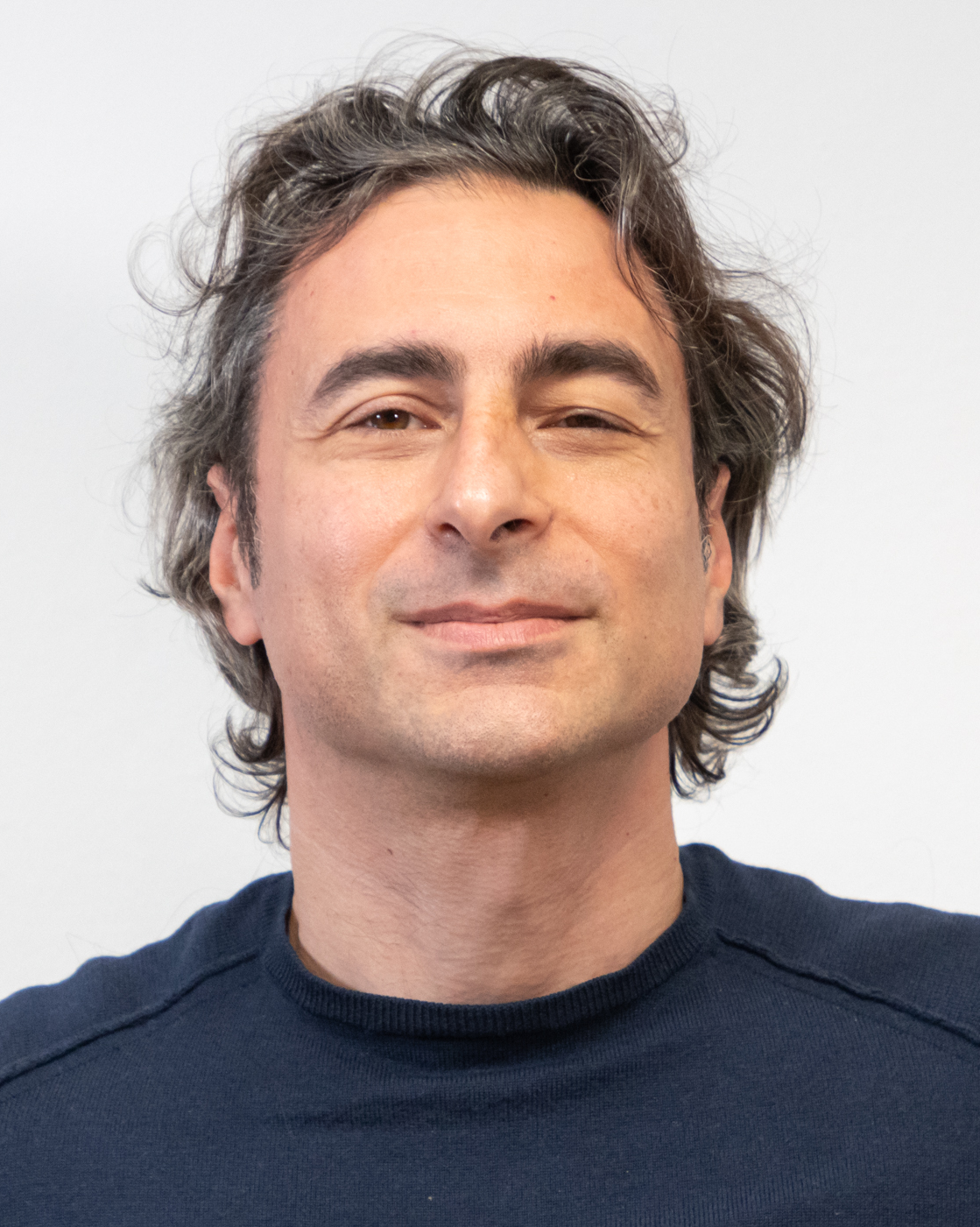}}]%
	{Michele Rossi} (SM'13) is a Professor of Wireless Networks in the Department of Information Engineering (DEI) at the University of Padova (UNIPD), Italy, where is the head of the Master's Degree in ICT for internet and Multimedia (\url{http://mime.dei.unipd.it/}). He also teaches Human Data Analysis at the Data Science Master's degree at the Department of Mathematics (DM) at UNIPD (\url{https://datascience.math.unipd.it/}). Since 2017, he has been the Director of the DEI/IEEE Summer School of Information Engineering (\url{http://ssie.dei.unipd.it/}). His research interests lie broadly in wireless sensing systems, green mobile networks, edge and wearable computing. Over the years, he has been involved in several EU projects on wireless sensing and IoT and has collaborated with major companies such as Ericsson, DOCOMO, Samsung and INTEL. His research is currently supported by the European Commission through the H2020 projects MINTS (no. 861222) on ``mmWave networking and sensing'' and GREENEDGE (no. 953775) on ``green edge computing for mobile networks'' (project coordinator). Dr. Rossi has been the recipient of seven best paper awards from the IEEE and currently serves on the Editorial Boards of the IEEE Transactions on Mobile Computing, and of the Open Journal of the Communications Society.
\end{IEEEbiography}

\end{document}